\documentstyle[12pt]{article}
\input amssym.def       
\input amssym.tex

\def\double{\Bbb}

\def\cc{{\double C}}     
       
\def\rr{{\double R}}     
\def\zz{{\double Z}}     
\def\qqq{{\double Q}}

\def\mm{{{\cal M}}}
    
\def\aa{{\cal A}}

\def\ep{{\cal E}}
\def\pe{{\petit{\rm E}}}

\def\ps{{\petit{\rm S}}}
\def\sdi{{{\cal D}(\ep)}}
\def\enep{{\hbox{End}(\ep)}}
\def\penep{{\petit{\rm End(\ep)}}}

\def\cb{{{\cal C}(\mm)}}

\def\sz{{\hbox{End}_{Cl}(\ep)}}
\def\psz{{\petit{\rm End}_{Cl}(\ep)}}

\def\di{\,\hbox{\rm D}}
\def\de{\,\hbox{\rm d}}

\def\dit{\tilde{\,\hbox{\rm D}}}
\def\pdit{\petit{\tilde{\,\hbox{\rm D}}}}

\def\dipi{\,\hbox{\rm D}_\phi}
\def\naep{\,\hbox{$\nabla^{\cal E}$}\,}
\def\pnaep{{\petit{{\nabla}^{\cal E}}}}

\def\nat{\,\hbox{$\tilde{\nabla}^{\cal E}$}\,}
\def\nap{{\,\hbox{${\tilde\nabla}'^{\cal E}$}\,}}

\def\pnap{{\petit{{{\tilde\nabla}'^{\cal E}}}}}

\def\sc{{\nabla\!\!\!\!\nabla}^\ep}

\def\pdi{{\petit{\rm D}}}

\def\ot{\otimes}
\def\op{\oplus}

\def\mapright#1{\smash{\mathop{\longrightarrow}\limits^{#1}}}

\def\bb{\begin{eqnarray}}
\def\ee{\end{eqnarray}}
\def\eee{\nonumber\end{eqnarray}}

\begin{document}

\hsize 17truecm
\vsize 24truecm
\font\twelve=cmbx10 at 13pt
\font\eightrm=cmr8
\def\petit{\def\rm{\fam0\eightrm}}
\baselineskip 18pt

\begin{titlepage}

\centerline{\twelve HUMBOLDT-UNIVERSIT\"AT ZU BERLIN}
\centerline{\twelve Unter den Linden 6}
\centerline{\twelve D-10099 Berlin}
\vskip 2truecm

\centerline{\twelve  The Einstein-Hilbert-Yang-Mills-Higgs
                                 Action}
\centerline{\twelve  and} 
\centerline{\twelve  the Dirac-Yukawa Operator}

\vskip .5truecm

\centerline{by}
\begin{center}
{\it J\"urgen Tolksdorf}
\footnote{Tolkdorf@mathematik.hu-berlin.de}\\
\end{center}

\vskip 1truecm
\leftskip=1cm
\rightskip=1cm

\begin{abstract}

In this article we show in some detail how the full action 
functional of the standard model of elementary particle 
physics can be described within the geometrical setting of 
generalized Dirac operators. We thereby introduce a new
model building kit for (a certain class of) gauge invariant 
theories which provides a unified geometrical description of 
Einstein's theory of gravity and Yang-Mills gauge theories on 
the "classical" level. Moreover, when the gauge symmetry is 
spontaneously broken, the Higgs sector as well has a natural
geometrical interpretation. It turns out that the Higgs field
is related to the gravitational potential.\\

Since the full action functional of the standard model is 
derived in one stroke, the appropriate parameters of the 
model have to satisfy certain relations similar to those in 
the Connes-Lott approach. Likewise, this may yield some 
phenomenological consequences, which is illustrated by
using the gauge group of the standard model in the case of 
$N-$generations of leptons and quarks.

\end{abstract}  

\vskip 2.5truecm

\begin{tabbing}
PACS-92\quad\=
04.50 Unified field theories and other theories of 
gravitation\\
\>11.15 Gauge field theories\\
MSC-91\>81E13 Yang-Mills and other gauge theories\\
\>83E99 General relativity, unified field theories
\end{tabbing}

\smallskip

%\noindent arch-ive/9612149

\end{titlepage}

% Beginn des Artikels

\section{Introduction}

In this article we propose a certain model building kit
which permits derivation of the action functional of the 
standard model of elementary particle physics with gravity 
including in terms of generalized Dirac operators. For this 
we introduce the following geometrical data:
\bb
({\rm G}, \rho, \di),
\ee
where G denotes a real compact semi-simple
Lie group, $\rho$ its unitary representation on a 
N-dimensional hermitian vector space $V$ and $\di$
denotes a Dirac operator acting on sections into a Clifford
module bundle $\ep\! :=\! {\rm S}\!\ot\!{\rm E}$, such that
${\rm E}\!:=\!{\cal P}\!\times_\rho\! V$. Here, S denotes
the spinor bundle over a closed compact orientable 
Riemannian spin manifold $(\mm, g)$ without boundary and 
of even dimension $(2n\!>\!2)$; ${\cal P}$ is a 
G-principal bundle over $\mm$.

Having given the geometrical setting we propose the
following functional
\bb
{\cal I}_{\rm D}:= 
(\psi, i\di\psi)_{\Gamma(\ep)}+{\rm res}_\zeta(\di^{-2n+2}),
\ee
Here, res denotes the Wodzicki residue; $\zeta$ is an element 
of the commutant defined by $({\rm G}, \rho)$, satisfying
$[\di,\zeta]\!=\![\chi, \zeta]\!=\!0,\,\zeta\! >\! 0$, with
$\chi$ the involution operator on $\ep$ and 
$(\; ,\;)_{\Gamma(\ep)}$ denotes the induced 
hermitian product on the $C^\infty(\mm)-$module
$\Gamma(\ep)$ of sections $\psi$ into $\ep$. In our frame 
this functional serves as a "general action" functional.

Using this frame the main result of this paper may be
summarized in the\\

\noindent
{\bf Theorem:}\\ 
{\it There exists a natural generalization (1) of the 
Dirac-Yukawa operator of the standard model such that 
the functional (2) is proportional to the full action of 
the standard model with gravity including}.\\

As a consequence we obtain certain relations between 
the parameters involved. Especially, the mass of the
Higgs field $m_h$ is a function of the fermion masses. 
In the most general case the range for the electroweak angle 
$\theta_w$ reads: $$0.25\leq\sin^2\theta_w\leq0.45\,.$$
In the case where all irreducible subspaces of the
fermion representation are equally weighted we obtain the 
GUT preferred relations
\bb
\sin^2\theta_w &=& 3/8,\nonumber\cr
g_{(3)} &=& g_{(2)},
\ee
where, respectively, $g_{(3)}\,\hbox{and}\,g_{(2)}$ are the 
strong and weak coupling constants. Of course, all derived 
relations are expected to be scale dependent and thus the 
corresponding renormalization flux must be carefully taken 
into account. We stress that the corresponding hypercharges
of the particles involved in the model are fixed when the
electrical charges are assumed to be known.

The model building kit as defined by the generalized Dirac 
operator (1) and the universal action (2) is motivated by the
assumption that the basic objects in nature are fermions
and that their dynamics is described by Dirac's equation.
Then, the appropriate dynamics of the various fields
involved in the definition of the fermionic interactions
should be a consequence of the latter and not independent
thereof. Mathematically, this may be rephrased as follows: 
when fermions are geometrically described by sections into a 
twisted spinor bundle, the most natural operator acting on 
those objects is a Dirac operator. A Dirac operator, however, is
but a Clifford superconnection (i.e. a certain generalization
of a connection) defined by (homomorphism-valued) 
differential forms of {\it various} degrees. Then, fixing the
admissible fermionic interactions geometrically means to
fix the admissible differential forms defining a certain
superconnection and thus a certain Dirac operator. The idea
of the kit, proposed here, is that the functionals leading to
the field equations of the differential forms defining the 
Dirac operator are not arbitrary but determined by this 
operator. As it turns out, this idea not only permits a 
geometrical understanding of the Higgs action but also a new 
geometrical interpretation of the Einstein-Hilbert and 
Yang-Mills functional (${\cal I}_{EH},\,{\cal I}_{YM}$). 
Indeed, from this point of view the former occures - in a 
sense - as a natural "companion" of the latter and both are 
"consequences" of the fermionic interaction. In this 
(classical) description of particle physics the a priori 
assumption of a flat spacetime seems artificial. To make this 
more clear, let us assume we consider a "free" fermion. From a 
geometrical viewpoint, such an object may be considered as a 
section into a twisted spinor bundle, where now the twisting 
part is  assumed to posess a trivial connection. Nevertheless, 
the Clifford module bundle may be non-flat. Indeed, the 
fermion carries energy and thereby produces a non-trivial 
gravitational field. Hence, the corresponding spin connection 
is non-trivial in general. Of course, under "normal" conditions
the energy of an elementary particle is so weak that the 
appropriate gravitational field cannot be measured. From our
point of view this simply means that in the particular case
at hand the energy-momentum tensor - defined by the Dirac 
action - is ignorable with respect to any inertial frame and, 
therefore, spacetime becomes {\it approximately} flat. 
Though this is usually accepted by physicists the point
here is that the functional ${\cal I}_{\rm D}$ fixes the 
energy-momentum tensor. Consequently, in our approach a 
vanishing Einstein tensor a priori makes no sense. This
can strictly hold iff there are no fermions in the world
(what ever such a world looks like!). 
Next, let us consider the case when the fermion becomes 
massive. Then, the interaction with the Higgs field must be 
taken into account. As it turns out, geometrically, the Higgs 
field defines a certain connection on the Clifford module 
bundle where the fermions live in. The corresponding 
curvature is non-zero, even in the case when the Higgs field 
represents a (classical) non-trivial vaccum. Of course, since 
the energy-momentum of the Higgs field is non-zero, 
spacetime must be curved as well. Note that this holds true 
even in the case when the world "sits in the (classical) 
vaccum". We stress that in our scheme the Higgs field seems 
intimately related to gravity. Of course, we are only 
considering classical field theory and one may object that on the level
of elementary particles quantum theory has to be taken into
account and then gravity may look completely different than 
described by Einstein's equation. However, as a "first step" 
towards a real understanding of the interplay between
gravity and particle physics it might be useful to have a 
unified geometrical description of all interactions
on the level of the classical field equations known so far.\\

Mathematically, it is evident that the action (2) is gauge
invariant. Hence, our kit provides a general scheme for 
building (a certain class of) gauge invariant theories. It 
therefore might be worth remembering the input of a 
general Yang-Mills model and to compare both building kits,
correspondingly,. Here, we adopt the notation as given in 
\cite{IS}.\\

 The input of a {\it general Yang-Mills model} consists of the 
following data:
\begin{enumerate} 
\item a finite dimensional real, compact Lie group G,
\item a unitary representation 
$\rho_f=\rho_{\rm L}\op\rho_{\rm R}$ on the $\zz_2-$ 
graded hermitian vector space ${\rm V}\!_f=
{\rm V}_{\rm L}\op {\rm V}_{\rm R}$ 
of the left and right handed fermions,
\item a finite set of positive constants $\{g_{(k)}\}$
(the gauge coup\-ling con\-stants), pa\-ra\-me\-triz\-ing 
the general Killing form of the Lie algebra ${\cal G}$ of G; the 
number of these con\-stants is defined as the number of 
simple components of ${\cal G}$, including u(1) factors.
\item a unitary representation $\rho_h$ on a hermitian
vector space ${\rm V}\!_h$ of the Higgs field,
\item a G-invariant polynomial (Higgs potential)
${\rm V}\!_h\mapright{V}\rr$ of order four, which is bounded
from below,
\item one complex constant (the Yukawa coupling constant)
$g_y$ for every one dimensional invariant subspace in the
decomposition of the representation
$$({\rm V}^*\!_L\ot{\rm V}\!_R\ot {\rm V}\!_h)\op
({\rm V}^*\!_L\ot{\rm V}\!_R\ot {\rm V}^*\!_h),$$
\item an action functional
$${\cal I} :=
{\cal I}_{EH} + {\cal I}_{Dirac} + {\cal I}_{Yukawa} + 
{\cal I}_{YM} + {\cal I}_{Higgs}\,.$$
\end{enumerate}
Usually, it is assumed that spacetime is flat, so that 
${\cal I}_{EH}$ is ignored. In particular, the {\it standard 
model} is defined by
\bb
{\rm G}:= {\rm SU}(3)\times{\rm SU}(2)\times{\rm U}(1)
\ee
with three coupling constants $(g_{(3)}, g_{(2)}, g_{(1)})$;
\bb
{\rm V}_{\rm L} &:=&
\bigoplus_1^3\left[(1, 2, -1/2)\op(3, 2, 1/6)\right],\\[.5em]
{\rm V}_{\rm R} &:=&
\bigoplus_1^3\left[(1, 1, -1)\op(3, 1, -1/3)
\op(3, 1, 2/3)\right],
\ee
where $(n_3, n_2, n_1)$ denote the tensor product,
respectively, of an $n_3$ dimensional representation of SU(3), 
an $n_2$ dimensional representation of SU(2) and a one 
dimensional representation of U(1) with "hypercharge" $y$: 
$\rho(e^{i\theta}):=e^{iy\theta},\;y\in\qqq,\,
\theta\in[0,2\pi[$; 
\bb
V:=\lambda(\varphi\varphi^*)^2-
\mu^2\varphi\varphi^*\,,
\ee
with $\lambda, \mu >0$. There are 27 Yukawa coupling
constants which, however, are not all independent. In fact,
the standard model can be parametrized by 18 constants,
c.f. \cite{Na}.\\

The full action functional (i.e. with gravity including) in the 
usual description of the standard model (N=3) 
reads
\bb
{\cal I} &:=&
\hbox{$\frac{1}{16\pi\,G}$}
\int_\mm\!\ast r_\mm\\[.5em]
&+& 
\int_\mm\!\ast\Big({\psi^{(l)}}^*\,i{\rm D}_{(l)}{\psi^{(l)}}
+ {\psi^{(q)}}^*\,i{\rm D}_{(q)}{\psi^{(q)}}\Big)\\[.5em] 
&-&
\int_\mm\!\ast\Big(\sum_{i,j=1}^{{\rm N}}
({{\bf g}^{(l)}_y})_{ij}\,{\bar\psi}^{(l)}_{{\rm L}_i}\,
(\gamma_5\varphi)\,\psi^{(l)}_{{\rm R}_j} + 
({{\bf g}^{(q)}_y})_{ij}\,{\bar\psi}^{(u)}_{{\rm L}_i}\,
(\gamma_5\varphi)\,\psi^{(u)}_{{\rm R}_j}
+\cr
& &\phantom{\int_\mm\!\ast\Big(\sum_{i,j=1}^{{\rm N}}}
+ ({{\bf g}'^{(q)}_y})_{ij}\,{\bar\psi}^{(d')}_{{\rm L}_i}\,
(\gamma_5\,{\hbox{\boldmath$\epsilon$\unboldmath}}
{\bar\varphi})\,\psi^{(d')}_{{\rm R}_j}\,\Big)\,+\,
\hbox{comp. conj.}\\[.5em]
&+&
\hbox{$\frac{1}{g^2_{(3)}}$}
\int_\mm\!{\rm tr}({\bf C}\wedge\ast{\bf C})
+ \hbox{$\frac{1}{g^2_{(2)}}$}
\int_\mm\!{\rm tr}({\bf W}\wedge\ast{\bf W})
+ \hbox{$\frac{1}{2g^2_{(1)}}$}
\int_\mm\!B\wedge\ast B\\[.5em]
&+& 
\int_\mm\!{\rm tr}
(\,(\nabla\varphi)^*\wedge\ast(\nabla\varphi)\,) + 
\int_\mm\!\ast V\\[.5em]
&\equiv&
{\cal I}_{EH} + {\cal I}_{Dirac} + {\cal I}_{Yukawa} + 
{\cal I}_{YM} + {\cal I}_{Higgs}.
\ee
The traces in the definition of the Yang-Mills 
action (10) are taken with respect to the corresponding 
fundamental representations of SU(3) and SU(2). Note that 
$\mm$ denotes a Riemannian manifold, which explains the 
occurence of the apparently wrong relative sign in front of 
the Higgs potential and the occurence of $\gamma_5$ in the 
Yukawa coupling term (9).\\

In contrast to a general Yang-Mills model our proposed kit 
(1) - (2) has the following input:
\begin{enumerate}
\item a finite dimensional real, compact Lie group G,
\item a unitary representation 
$\rho =\rho_{\rm L}\op\rho_{\rm R}$ on the $\zz_2-$ 
graded hermitian vector space ${\rm V} =
{\rm V}_{\rm L}\op {\rm V}_{\rm R}$ 
of the left and right handed fermions,
\item a Dirac operator $\di$, 
\item the general functional: ${\cal I}_{\rm D}$.
\end{enumerate}
Like in the "non-commutative approach" as
introduced by Connes and Lott (c.f. \cite{CL}, \cite{Co1}), 
the Higgs representation is not arbitrary but has to lie within
the fermionic representation, symbolically: 
$\rho_h\subset\rho_f$. In fact, this has significant 
consequences with respect to the relations between the
various parameters involved in the model. Hence it
might not be come as a surprise that there are certain
similarities between the Connes-Lott model and the 
model introduced here in this respect (see below). Of course, 
the mathematical background is quite different. We mention 
that with respect to our physical interpretation it is quite 
natural  that all fields involved in the model carry the same 
representation.

Concerning the {\it standard model}, the Dirac operator 
$\di\equiv{\dit}_\phi$ is defined by a generalization of the 
{\it Dirac-Yukawa operator} $\dipi$. It is well-known that the 
Yukawa coupling (9) together with the (standard) Dirac 
operator $\di$ defining (8) can be considered as a new Dirac 
operator $\dipi$ - the Dirac-Yukawa operator. The main 
feature, then, is that this Dirac operator is a "non-standard" 
Dirac operator (i.e. not associated with a Clifford connection, 
see below). In fact, such Dirac operators will play a key role 
in our geometrical description of the standard model. 
Correspondingly, we shall discuss those operators in some
detail in the {\bf first part} of our paper, which is totally 
concerned with the mathematical frame of our model.
Though the larger portion of part 1 is actually not new
and may be found in much more detail, e.g., in 
\cite{BGV} we nevertheless summarize the basic mathematical 
notions in order for our paper to be self-contained and 
to permit full understanding of the issue 
also for those who are not familiar with the notion of 
non-standard Dirac operators. Also, we have emphazised the 
relations between non-standard Dirac operators and 
connections on a (general) Clifford module bundle. This is of 
technical significance and, moreover, explains how the Higgs 
field yields a certain connection on a Clifford module bundle.
In {\bf part two} we introduce a certain generalization of the 
Dirac-Yukawa operator and prove our main theorem. Moreover, 
we also investigate the "phenomenological" concequences of 
our scheme with respect to the standard model. Finally, we 
mention some similarities to the Connes-Lott model 
(c.f. \cite{CL}, \cite{Co1} and, concerning the new approach, 
\cite{CC1}, \cite{IKS1}). We conclude this paper with an 
outlook.

Before we start to describe the mathematical frame of our 
model building kit, however, some remarks concerning its 
similarity to the Connes-Lott approach to the standard model 
seem appropriate. Obviously, our notion of a generalized
Dirac operator, as defined by (1), is similar to Connes' notion 
of a "spectral triple". Needless to say that the 
latter notion is more profound, mathematically, since it 
offers the possibility of "new mathematics", like
Connes' non-commutative geometry, c.f. \cite{Co1}. Also,
the idea to {\it derive} specific functionals, such as in the
case of the standard model of particle physics, from a 
"universal functional" must go back to Connes. For example, 
in the Connes-Lott approach to the standard model the Dixmier
trace serves as the general action functional, c.f. \cite{Co1}, 
\cite{CL}. In a sense this trace can be considered as a 
special case of the more general Wodzicki residue. 
Unfortunately, using the Dixmier trace as the universal action 
functional it seems hard to derive both the Einstein-Hilbert 
and the Yang-Mills (-Higgs) action in one stroke (for the pure 
EH-functional see \cite{CFG}). Infact, in the Connes-Lott
description of the standard model the geometric information
contained in the Dirac operator is lost. But as Connes has 
remarked the Wodzicki residue of $\di^{-2n+2}$ with 
$\di$ a Standard Dirac operator (see below) becomes 
proportional to the Einstein-Hilbert action of gravity, 
c.f. \cite{Co2}. However, this time the geometrical information 
contained in the Yang-Mills potential is lost. This follows 
immediately from $\di$ being a Standard Dirac operator (see 
below). As a natural question one may ask whether it is 
possible to derive the full action functional (EHYMH-action) 
by considering "non-standard" Dirac-operators.
In \cite{AT2} this has been investigated and affirmatively 
answered in the case of the Einstein-Hilbert-Yang-Mills 
functional . In the case of the full action of the standard 
model and with gravity included the above question was 
investigated in \cite{AT3}. However, in this work there is a 
mistake. Infact, the definition of the Dirac-Yukawa operator is 
wrong and as a consequence the derived functional does not
coincide with the action functional of the standard model. As 
we shall show in the paper at hand, however, by using the 
right definition of the Dirac-Yukawa operator the properly 
corrected generalized Dirac-Yukawa operator proposed in 
\cite{AT3} is but a special case of the generalized 
Dirac-Yukawa operator introduced in part two of 
our paper and which infact gives rise to the full action of the 
standard model. Though the basic idea of our kit is already
introduced in \cite{AT3}, and which indeed affirmatively 
answers the above mentioned question, the scheme in 
\cite{AT3}, however, is still not general enough to discuss  
physical implications of the proposed model building kit. 
This is because in \cite{AT3} neither the considered Dirac 
operator - even if properly corrected - nor the proposed 
universal action functional is general enough. Both has been 
remedied in this article.

\section{Part 1: The mathematical frame}
The geometrical setting which we propose in order to 
describe gauge theories is that of a Clifford module bundle 
$(\ep, c)$ over a Riemannian manifold $(\mm, g)$ 
of even dimension. Within this setting there exists a 
distinguished class of operators called generalized Dirac 
operators\footnote{In the following, the term "generalized" 
simply means "non-standard", though by a 
generalized Dirac operator we actually mean the triple 
$({\rm G}, \rho, \di)$, with $\di$ a (non-standard) Dirac 
operator.}. We therefore start with a brief review on the 
notion of Clifford modules and generalized Dirac operators.
More details of this issue can be found, e.g., in \cite{BGV}.
Afterwards we shall discuss in some length how a given
generalized Dirac operator (1) determines a particular
functional ${\cal I}_\pdi(\naep, \psi)$, using (2).\\

To get started, let us denote by $(\mm, g)$ a smooth, 
closed compact {\it Riemannian} (spin) manifold without 
boundary and of even dimension: 
$\hbox{dim}(\mm)\!\equiv\! m\! :=\! 2n (>2)$. Moreover, let
$\ep\!:=\!\ep^+\!\op\ep^-$ be (the total space of) a 
$\zz_2-$graded hermitian vector bundle 
$\ep\mapright{\pi}\mm$ over $\mm$. The corresponding 
hermitian product on $\ep$ is indicated by $(\, ,\,)_\ep$. 
If $\Gamma(\ep)$ denotes the $C^\infty(\mm)-$module of 
smooth sections into $\ep$, then the induced hermitan  
product 
$(\, ,\,)_{\Gamma(\ep)}$ on $\Gamma(\ep)$ is given by:
$(\, ,\,)_{\Gamma(\ep)}\, :=\int_\mm\ast(\, ,\,)_\ep\,$. Here,
$"\!\ast\!"$ means the Hodge map regarding the
Riemannian metric $g$ on $\mm$.\\

\noindent
{\bf Definition 1:} A {\sl generalized Dirac operator} $\di$ 
is any odd first order differential operator 
acting on sections $\psi\in\Gamma(\ep)$:
\bb
\di:\Gamma(\ep^\pm)\rightarrow\Gamma(\ep^\mp),
\ee
so that $\di^2$ is a {\it generalized Laplacian}. I.e., there 
exists a connection: 
$\Gamma(\ep)\mapright{{\hat\nabla}^\ep}
\Gamma(T^*\mm\ot\ep)$
on the vector bundle $\ep$ and an endomorphism 
${\cal F}\in\Gamma(\enep)$,
both uniquely defined by $\di$, such that
\bb
\di^2 = \triangle^{\!\!{\hat\nabla}^\ep} + {\cal F}.
\ee
Here, $\triangle^{\!\!{\hat\nabla}^\ep}\! :=\! 
-ev_g({\hat\nabla}^{T^*\!\mm\ot\ep}\,{\hat\nabla}^{\ep})$ 
denotes the horizontal (Bochner) Laplacian associated with 
the connection ${\hat\nabla}^\ep$, and $"ev_g"$ means the
evaluation map regarding the metric $g$.\\

\noindent
{\bf Remark 1:} $\di$ exists on $\ep$ iff $\ep$ denotes a 
{\it Clifford module} bundle over $(\mm, g)$. I.e. there exists 
a graded (left) action on $\ep$
\bb
c: \cb\times\ep\rightarrow\ep
\ee
of the Clifford bundle $\cb\!\mapright{\tau}\!\mm$ 
associated with the metric $g\,$\footnote{Here, the 
fiber $\tau^{-1}(x)$ is isomorphic to the Clifford algebra 
generated by the elements $u,v\!\in\! T^*_x\mm$, using the 
relation: $uv + vu\! :=\! -2g_x(u,v),\,\forall\, x\!\in\!\mm$.}. 
This holds because in case that $\di$ denotes a generalized 
Dirac operator, this operator induces via
\bb
\cb\times\ep &\rightarrow &\ep\cr
(df, \psi) &\mapsto & c(df)\psi:= [\di, f]\psi, \quad\; 
f\in C^\infty(\mm)
\ee
a graded left action of the Clifford bundle on $\ep$, c.f.
\cite{BGV}. Conversely, if $(\ep, c)$ denotes a Clifford module 
bundle over $(\mm, g)$, then the well-known construction
\bb
\Gamma(\ep)\mapright{\pnaep}
\Gamma(T^*\mm\ot\ep)\hookrightarrow
\Gamma(\cb\ot\ep)\mapright{c}\Gamma(\ep)
\ee
defines an operator $\di_{\pnaep}\!:=\!c(\naep)$ satisfying
(13) and (14) for any connection $\naep$ on $\ep$. 
Moreover, it is easily checked that such a
defined operator also fulfils (16).\\

Hence, from now on $\ep\equiv(\ep, c)$ will always
denote a Clifford module bundle over $\mm\equiv(\mm, g)$. As
a consequence, the endomorphism bundle $\enep$  is also a
Clifford module and it follows that (c.f. \cite{BGV}) 
\bb
\enep\simeq\cb\ot\sz,
\ee
where $\sz$ denotes the algebra bundle of bundle 
endomorphisms of $\ep$ supercommuting with the action
of $\cb$, i.e.
\bb
\sz:=\{\sigma\in\enep\,|\,|\![c(a), \sigma]\!| = 0, \;\forall\,
a\in\cb\}.
\ee
If in addition $\mm$ is assumed to be a spin manifold we
have: $\cb_\cc\simeq{\rm End(S)}$ and correspondingly 
$\sz\simeq{\rm End(E)}$, where S denotes
the {\it spinor} bundle and E a vector 
bundle over $\mm$. In this case $\ep$ is called a 
{\sl twisted spinor bundle} over $\mm$ and we have 
(c.f. \cite{BGV}) 
\bb
\ep\simeq {\rm S}\ot {\rm E}.
\ee
We say that the Clifford module bundle $(\ep, c)$ has 
a {\sl twisting graduation} if $\sz$ posesses a (non-
trivial) $\zz_2-$graduation, cf. \cite{AT2}. Clearly, in
the case of (20) this is equivalent to saying that the
vector bundle ${\rm E}={\rm E_L}\op {\rm E_R}$ is 
$\zz_2-$graded, as well.\\
 
\noindent
{\bf Definition 2:} Let $\dit$ be a generalized Dirac operator 
on $\ep$. It is said to be compatible with the Clifford action 
$c$ if it satisfies the relation (16). Then, let
\bb
\sdi := \{ \dit\,|\, [\dit, f] = c(df), \;\forall\, 
f\in C^\infty(\mm)\}
\ee
be the set of all generalized Dirac operators on $\ep$ which 
are compatible with the Cifford action $c$. 
We have\footnote{Since the
bundle $\ep\! =\!\ep^+\!\op\ep^-\,$ is $\zz_2-$graded, 
so is the associated endomorphism bundle ${\rm End}(\ep)$. 
I.e. ${\rm End}(\ep)\! =\!
{\rm End}^+(\ep)\!\op\!{\rm End}^-(\ep)$, with
${\rm End}^+(\ep)\! :=\!{\rm End}(\ep^+)\!\op\!
{\rm End}(\ep^-)$ and
${\rm End}^-(\ep)\! :=\!{\rm Hom}(\ep^+,\ep^-)
\!\op\!{\rm Hom}(\ep^-,\ep^+)$.}
\bb
\sdi\simeq\Omega^o(\mm, \hbox{End}^-(\ep)).
\ee
Also, let us denote by
\bb
\aa(\ep):=\{\naep\;|\;
\Gamma(\ep)\mapright{\pnaep}
\Gamma(T^*{\!\mm}\ot\ep)\,\}
\ee
the set of all (even) connections on $\ep$.\\

As a consequence of (18), there exists a natural class of
connections - called {\sl Clifford connections} - on any 
Clifford module bundle $\ep$:
\bb
\aa_{Cl}(\ep):=\{\naep\in\aa(\ep)\;|\; [\naep, c(a)] = 
c(\nabla^{\petit{Cl}}a), \;\forall\,a\in\Gamma(\cb)\,\}\subset
\aa(\ep),
\ee
where $\nabla^{\petit{Cl}}$ is the induced Levi-Cevita
connection on $\cb$. Note, in the case of a twisted spinor 
bundle (20) any Clifford connection 
$\naep\!\in\!\aa_{Cl}(\ep)$ takes the form of a tensor 
product connection (c.f. \cite{BGV})
\bb
\naep \equiv \nabla^{{\petit{{\rm S}\ot{\rm E}}}} := 
\nabla^{\ps}\ot{\bf 1}_{\rm E} + 
{\bf 1}_\ps\ot\nabla^{\pe},
\ee
where, respectively, $\nabla^\ps$ denotes the spin
connection on S and $\nabla^\pe$ any connection on
the vector bundle E. In general we have 
\bb
\aa(\ep)\! &\simeq &\!
\Omega^1(\mm, {\hbox{End}}^+(\ep)), \nonumber\\[.3em]
\aa_{Cl}(\ep)\! &\simeq &\!
\Omega^1(\mm, \hbox{End}_{Cl}^+(\ep)),
\ee
where the latter isomorphism follows from (18). \\ 

\noindent
{\bf Remark 2:} Using the linear isomorphism:
\bb
\Lambda T^*\!\mm &\mapright{\bf c}&\cb \cr
e^{i_1}{\petit{\wedge}}e^{i_2}\cdots
{\petit{\wedge}}e^{i_k}&\mapsto&
e^{i_1}e^{i_2}\cdots e^{i_k}, \quad\forall\; 
0\leq k\leq m,
\ee
between the Grassmann- and the Clifford bundle,
where $\{e^i\}_{1\!\leq i\leq\! m}$ denotes a 
local orthonormal basis in $T^*\!\mm$ and 
Clifford multiplication is indicated by juxtaposition, the
Clifford action $c$ induces a linear mapping,
also denoted by $c$, 
\bb
\Omega^p(\mm, {\rm End}^\pm(\ep))&\mapright{c}&
\Omega^o(\mm, {\rm End}^\mp(\ep)) \cr
\alpha &\mapsto & c(\alpha).
\ee\\
\noindent
{\bf Lemma 1:} {\it In the case of $p=1$ the linear 
mapping (28) has a canonical right inverse defined by
\bb
\delta_\xi: \Omega^{p}(\mm, {\rm End}^\mp(\ep))&
\rightarrow&\Omega^{p+1}(\mm, {\rm End}^\pm(\ep))\cr
\alpha &\mapsto &\xi\wedge\alpha,
\ee 
where $\xi\in\Omega^1(\mm, {\rm End}^-(\ep))$ is locally
given by}\footnote{Throughout this paper we adopt 
Einstein's convention for summation.}
\bb
\xi := -\frac{1}{m}\,
g(e_a, e_b)\,e^a\!\ot c(e^b)\ot{\bf 1}_\psz.
\ee
The product in (29) simply means: 
$(\sigma\ot a)\!\wedge\!(\sigma'\ot a')\!:=
\!\sigma\!\wedge\!\sigma'\!\ot aa'$ for all homogeneous
elements $\sigma\ot a, \sigma'\ot a'\in
\Omega^*(\mm, \enep)\simeq
\Gamma(\Lambda T^*\!\mm\!\ot\!\enep)$; Again,
$\{e^a\}_{1\leq a\leq m}$ is a basis in 
$T^*\!\mm$ and $\{e_a\}_{1\leq a\leq m}$ its dual. Note
that this "wedge product" is not graded commutative and
that $\Omega^*(\mm, {\rm End}(\ep))$ is a bi-graded
algebra. The map (29) is even with respect to the total
grading.\\

\noindent
{\sl Proof:} Obviously, this follows by construction.\\

\noindent
{\bf Remark 3:} The form $\xi$, locally defined by (30), can 
be characterized via
\bb
\nabla^{T^*\!\mm\ot\penep}\xi &\equiv& 0, 
\quad\forall\,\naep\!\in\aa_{Cl}(\ep),\cr
c(\xi) &=& {\bf 1}_\ep.
\ee\\
\noindent
{\bf Lemma 2:} {\it Let $c$ be the linear mapping (28) 
restricted to $\Omega^1(\mm, {\rm End}^+(\ep))$. 
Then we have}
\bb
\sdi\simeq\aa(\ep)/\,{\rm ker}(c).
\ee\\
\noindent
{\sl Proof:} For $p:=\delta_\xi c:
\Omega^1(\mm, {\rm End}^+(\ep))
\rightarrow\Omega^1(\mm, {\rm End}^+(\ep))$ we get
$p^2 = p$ and thus
\bb
\Omega^1(\mm, {\rm End}^+(\ep))={\rm im}(p)\op{\rm im}(q),
\ee
with $q:={\bf 1} - p$. Restricting the linear mapping (28) to 
$\Omega^1(\mm, {\rm End}^+(\ep))$ yields the identities
\bb
p c &\equiv& p, \cr
c p &\equiv& c.
\ee
Since $\delta_\xi$ is a right inverse of $c$ we have: 
${\rm ker}(c)\subset{\rm im}(q)$. Moreover,  
for all $\alpha\!\in\!{\rm im}(p)$ with $c(\alpha)=0$
(34) implies $\alpha\equiv 0$. Hence
\bb
{\rm ker}(c) = {\rm im}(q).
\ee 
The statement follows from
\bb
\aa(\ep)\simeq\Omega^1(\mm, {\rm End}^+(\ep))\mapright{c}
\Omega^o(\mm, {\rm End}^-(\ep))\simeq\sdi
\ee
and we are done.\\
Note that actually we have shown that the sequence:
$$0\rightarrow{\rm ker}(c)\rightarrow\Omega^1(\mm,{\rm End}^+\ep)
\,\mapright{c}\,\Omega^o(\mm,{\rm End}^-\ep)\rightarrow 0$$ 
is exact and splits. Also, in this case the kernel of the
mapping (28) becomes explicit.\\

\noindent
{\bf Definition 3:} Two connections 
$\naep, \nat\!\in\!\aa(\ep)$ are defined to be equivalent iff 
\bb
\naep - \nat \in {\rm ker}(c).
\ee
By the preceding Lemma this is equivalent to
\bb
\naep &\sim &\nat\Longleftrightarrow\nat =
\naep +\omega,\cr
\omega &\in & {\rm im}(q)
\subset\Omega^1(\mm, {\rm End}^+(\ep)).
\ee\\
Therefore, any Dirac operator $\di\!\in\!\sdi$ is
uniquely associated with an equivalence class of connections
$[\naep]$ on $\ep$ so that $\di_{\pnaep}\!=\!\di$.\\

\noindent
{\bf Remark 4:} Let $\dit\!\in\!\sdi$ be a given Dirac 
operator on $\ep$. Then,
\bb
\nat := \naep + \delta_\xi(\dit - \di_\pnaep)
\ee
defines a connection on $\ep$, so that
\bb
\di_{{\tilde{\nabla}^\ep}} = \dit,
\ee
where $\naep\!\in\!\aa(\ep)$ denotes any connection 
on $\ep$. Clearly, this ambiguity simply reflects that
$\sdi$ is an affine space and thus a given Dirac operator 
$\dit\!\in\!\sdi$ may be decomposed in infinitely many ways 
like
\bb
\dit = \di + \Phi_{\rm D}
\ee
with 
$\Phi_{\petit{\di}}\!:=\dit - \di\!\in\!
\Omega^o(\mm,{\hbox{End}}^-(\ep))$.\\

\noindent
{\bf Definition 4:}
We call a Dirac operator $\di\!\in\!\sdi$ a 
{\sl Standard Dirac operator} (SDO) if there is a Clifford
connection $\naep\!\in\!\aa_{Cl}(\ep)$, so that
\bb
\di = \di_\pnaep.
\ee\\
\noindent
{\bf Lemma 3:} {\it Let $\naep, \nat\in\aa_{Cl}(\ep)$ be 
Clifford connections on the Clifford module bundle $\ep$ 
with $\naep\sim\nat$. Then we have:} $\naep\equiv\nat$.\\

\noindent
{\sl Proof:}
Using (38), $\naep, \nat\!\in\!\aa_{Cl}(\ep)$ implies that 
there exists a $\omega\!:=\!\naep\! -\! \nat\! =\! 
{\bf 1}_{Cl}\ot A$ 
with $A\!\in\!\Omega^1(\mm, {\rm End}^+_{Cl}(\ep))$. By
assumption, we have 
\bb
c(\omega) &=& c(e^\mu)\ot A_\mu\cr
&=& 0,
\ee
where $\{e^\mu\}_{1\leq\mu\leq m}$ is a local orthonormal 
frame. Hence, $A_\mu\! =\! 0,\forall\,\mu\! =\! 1,\cdots,m$ 
which proves the lemma.\\

We therefore have shown that the class of connections
defining a SDO on $\ep$ admits a canonical representative.
In what follows we shall always denote by $\di\!\in\!\sdi$ a 
SDO and by $\naep\!\in\!\aa_{Cl}(\ep)$ the appropriate 
Clifford connection, so that $\di\! =\!\di_{\nabla}$. In
contrast, by $\dit\!\in\!\sdi$ and $\nat\!\in\!\aa(\ep)$, 
respectively, we denote an arbitrary Dirac operator and 
connection on $\ep$.\\

Let $\dit\!\in\!\sdi$ be an arbitrary Dirac operator on the 
Clifford module bundle $\ep$ and let  
\bb
\nat := \naep + \delta_\xi(\dit - D) 
\ee
Then, in \cite{AT2} it is shown that (see also \cite{Bi})
\bb
\dit^2 = \triangle^{\!\!{\hat\nabla}^\ep} +
{\cal F}^{{\tilde\nabla}^\ep}.
\ee
Here, respectively, the connection 
${\hat\nabla}\!\in\!\aa(\ep)$ and the endomorphism
${\cal F}^{{\tilde\nabla}^\ep}\!\in\!\Gamma(\enep)$ are
defined by
\bb
{\hat\nabla}^\ep &:=& \nat + 
\omega_{{\tilde\nabla}^\ep}\;\hbox{and}\\[.5em] 
{\cal F}^{{\tilde\nabla}^\ep} &:=&
c\left({\nat}^2\right) + 
{\rm ev}_g\left({\tilde\nabla}^{T^*\!\mm\ot\penep}
\omega_{{\tilde\nabla}^\ep} + 
\omega_{{\tilde\nabla}^\ep}^2\,\right),
\ee
where the one form 
$\omega_{{\tilde\nabla}^\ep}\!\in\!
\Omega^1(\mm, {\rm End}^+(\ep))$ is locally given by
\bb
\omega_{{\tilde\nabla}^\ep} := -\frac{1}{2}g(e_\mu,e_\nu)\,
 e^\mu\ot c(e^\lambda)\left([\nat\!\!_{\lambda}, c(e^\nu)] +
\Gamma^{\nu}_{\;\sigma\lambda}\,c(e^\sigma)\right).
\ee
The $\Gamma$'s denote the Christoffel symbols defined by
the metric $g$.\\

\noindent
{\bf Lemma 4:} {\it The endomorphism 
${\cal F}^{{\tilde\nabla}^\ep}$ defined in (47) is independent 
of the representative  $\nat\!\in\!\aa(\ep)$ of the class of 
connections defining the Dirac operator} $\dit$.\\

\noindent
{\sl Proof:} To prove this lemma we introduce the affine 
mapping
\bb
\varpi: \aa(\ep) &\rightarrow &\aa(\ep)\cr
\nat &\mapsto &\nat + \omega_{{\tilde\nabla}^\ep}
\ee
on $\aa(\ep)$ and show that this mapping is well-defined on
$\aa(\ep)/\,{\rm ker}(c)$. Let $[\nat]$ be the equivalence
class of connections defining the given Dirac operator
$\dit\!\in\!\sdi$ and denote by $\nat,\!\nap\!\in\![\nat]$
two representatives of this class. Hence, 
$\alpha\!:=\!\nat\!-\!\nap\!\in\!{\rm ker}(c)$ and with
respect to a local orthonormal frame we obtain
\bb
{\hat\nabla}^{\cal E} - {\hat{\nabla}}'^{\cal E} &=& \alpha -
\frac{1}{2}\delta_{\mu\nu}\,e^\mu\ot c(e^\lambda)\,
[i_\lambda\alpha, c(e^\nu)]\\[.5em]
&=&
\frac{1}{2}\delta_{\mu\nu}\,e^\mu\ot 
[c(\alpha), c(e^\nu)]_+\\[.5em]
&=& 0,
\ee 
where $i_\mu$ is the inner derivative with respect to the
local vector field $e_\mu$ and $[\; ,\;]_+$ means the 
anti-commutator. Consequently, the map: 
$\dit\mapsto\triangle^{\!\!{\hat\nabla}^\ep}$, with
${\hat\nabla}^\ep$ given by (46), is well-defined. Hence
the endomorphism ${\cal F}^{{\tilde\nabla}^\ep}\! =\! \dit^2 -
\triangle^{\!\!{\hat\nabla}^\ep}$ only depends on the
Dirac operator $\dit\!\in\!\sdi$ which proves the lemma.\\

\noindent
{\bf Corollary 1:} {\it The endomorphism 
${\cal F}^{{\tilde\nabla}^\ep}\!\in\!\Gamma(\enep)$ does not
depend on the decomposition (44). In particular, it does not
depend on the chosen Clifford connection} $\naep$.\\

\noindent
{\sl Proof:} By the preceding lemma 4 the proof is obvious.\\

\noindent
{\bf Remark 5:} The corresponding linear part of the affine 
map (49) has a non-trivial kernel; especially we get
\bb
\varpi|_{\aa_{Cl}(\ep)} = {\bf 1}_{\aa_{Cl}(\ep)}.
\ee  
In this case the decomposition of the square of the
appropriate SDO is but the usual {\sl Lichnerowicz
formula} and the endomorphism ${\cal F}$ takes its 
well-known form
\bb
{\cal F}^\pnaep &=& c[(\naep)^2]\\[.5em]
&=& \hbox{$\frac{1}{4}$}\,r_\mm\,{\bf 1}_\ep +
c(F^{\ep/{\rm S}}),
\ee 
where $r_\mm$ is the Ricci scalar curvature
on the base manifold $\mm$ and $F^{\ep/{\rm S}}\!:=
(\!\naep\!)^2 - (\nabla^{Cl})^2\!\ot\!{\bf 1}_{\psz}$ 
denotes the relative curvature on the Clifford module bundle
$\ep$. Since in this particular case the relative curvature 
only depends on the connection on the twisting part of the 
Clifford module bundle $\ep$, $F^{\ep/{\rm S}}$ is also 
called the {\it twisting curvature}.\\

We now turn to the notion of superconnections which
can be considered as a generalization of connections
on a $\zz_2-$graded vector bundle. As it is well-known
superconnections permit to generalize the one two one
correspondence between SDO and Clifford connections to
arbitrary Dirac operators and Clifford superconnections on a 
Clifford module bundle. Hence, there is no essential difference 
between talking about Dirac operators and Clifford 
superconnections. We therefore call into mind the following\\

\noindent
{\bf Definition 5:} 
A {\it superconnection} on a $\zz_2-$graded vector bundle
$\ep$ is any odd first order differential 
operator\footnote{Here, $\pm$ is understood with respect
to the {\it total} grading.}
\bb
\sc : \left[\Omega^*(\mm, \ep)\right]^\pm \rightarrow
\left[\Omega^*(\mm, \ep)\right]^\mp,
\ee
satisfying the generalized Leibniz rule
\bb
\sc(\lambda\wedge\alpha) = \de\lambda\wedge\alpha +
(-1)^{|\alpha|}\,\alpha\wedge\sc\alpha,
\ee
for all $\lambda\!\in\!\Omega^*(\mm)$ and 
$\alpha\!\in\!\Omega^*(\mm, \ep)$. If in addition $\ep$
denotes a Clifford module bundle and the superconnection
fulfils
\bb
[\sc, c(a)] = c(\nabla^{Cl} a), \quad\forall\, a\in\Gamma(\cb),
\ee
it is called a {\sl Clifford superconnection} (CSC), 
c.f. \cite{BGV}. In this case we have (c.f. loc. cit.)
\bb
\sc\mapsto\di_{\petit{\sc}} := c(\sc)
\ee
is one to one\footnote{Note, by abuse of notation we suppress 
the linear isomorphism (27).}. Note that because of the 
generalized Leibniz rule any superconnection locally takes 
the form
\bb
\sc &=& \de + \sum\limits^m_{k=0}\aa_{[k]},\cr
\aa_{[k]} &\in &\left[\Omega^k(\mm,\enep)\right]^-\,.
\ee
In particular, in the case of a CSC
\bb
\aa_{[1]} &=& \omega^{Cl}\ot{\bf 1}_\psz +
{\bf 1}_{Cl}\ot A \nonumber\\[.5em]
\aa_{[k]} &=& \sum\limits_{1\leq i_1<\cdots <i_k\leq m}
e^{i_1}\wedge\cdots\wedge e^{i_k}
\ot{\bf 1}_{Cl}\ot B_{i_1\cdots i_k}, \quad\forall\,
k = 0, 2,\cdots ,m\,
\ee
where $\omega^{Cl}$ denotes the induced
Levi-Civita form on the Clifford bundle ${\cal C}(\mm)$ and
$A, B\!\in\!\left[\Omega^*\left(\mm,\sz\right)\right]^-$.
Again, $\{e^i\}_{1\leq i\leq m}$ is a local orthonormal 
frame in $T^*\!\mm$. Moreover, if $\mm$ denotes a 
spin-manifold, then any CSC is of the form (c.f. \cite{BGV})
\bb
\sc = \nabla^{\rm S}\ot{\bf 1}_{\rm E} +
{\bf 1}_{\rm S}\ot{\nabla\!\!\!\!\nabla}^{\rm E}.
\ee
Clearly, the notion of a CSC completely parallels that of a
Clifford connection and coincides with the latter iff 
$B\!\equiv\!0$. However, in general a CSC is not just defined 
by an element of
$\Omega^1\!\left(\mm,{\rm End}^+_{Cl}(\ep)\right)$. This 
will be of crucial importance in what follows. Indeed, in 
\cite{AT2} it was shown how the combined 
Einstein-Hilbert-Yang-Mills (EHYM-) action functional can 
be derived using non-SDO's. Before we define a particular 
functional on $\sdi$ we still give another\\

\noindent
{\bf Remark 6:} Let $\dit\!\in\!\sdi$ be a Dirac operator 
on $\ep$ and $\sc$ the corresponding CSC. Then, we have
\bb
\nap &:=& \naep + \omega,\quad\hbox{with}\\[.5em]
\omega &:=& \omega_\phi + e^i\ot
\left(\sum\limits^{m-1}_{k=1}\;\frac{1}{k!}
\sum\limits^{m}_{i_1,\cdots, i_k=1}
c(e^{i_1})\cdots c(e^{i_k})\ot B_{i\, i_1\cdots i_k}\right),
\ee
where 
$\omega_\phi\!:=\!\delta_\xi\Phi,\;\Phi\!\equiv\!\aa_{[0]}
\!\in\!\Omega^o(\mm,{\rm End}^-(\ep))$, so that
\bb
\di_{\pnap} = \dit.
\ee
In particular, $\nap\sim\nat$, where the latter is defined
by (44). Note, the local decomposition of the form
$\omega\!\in\!\Omega^1(\mm,{\rm End}^+(\ep))$ may also
contain a degree $[k]=1$ form.\\

After summarizing the notion of
Clifford modules and generalized Dirac operators we have
also proved some lemmas, which permit an understanding
of the relations between Dirac operators and connections on 
a Clifford module bundle. The reason to clarify these relations 
mainly is motivated by the following\\

\noindent
{\bf Definition 6:}
Let $(\mm, g)$ be a closed, compact, orientable 
Riemannian manifold of even dimension ($m=2n>2$) and 
without boundary. Also, let $(\ep, c)$ be a Clifford module 
bundle over $\mm$ and let $\sdi$ be the affine space of all
(generalized) Dirac operators compatible with the Clifford
action $c$ on $\ep$. Then, we introduce the functional
\bb
{\rm res} : \sdi &\rightarrow & \cc \cr
\dit &\mapsto & {\rm res}(\dit^{-2n+2}).
\ee 
Here, ${\rm res}$ means the Wodzicki residue, which in this 
case takes the explicit form
(cf. \cite{Wo1}, \cite{Wo2}, \cite{Gu}, \cite{KW})
\bb
{\rm res}(\dit^{-2n+2}) = 
\hbox{$\frac{2}{\Gamma(n-1)}$}
\int_\mm\!
\ast{\rm tr}_\ep\left[\hbox{$\frac{1}{6}$}\,r_\mm - 
{\cal F}^{{\tilde\nabla}^\ep}\right]
\ee
and where the endomorphism 
${\cal F}^{{\tilde\nabla}^\ep}\!\in\!\Gamma(\enep)$
is given by (47).\\

Since there exists a connection $\nat\!\in\!\aa(\ep)$
for every $\dit\!\in\!\sdi$, so that 
$\dit\!=\!\dit_{{\tilde\nabla}^\ep}$, this functional may
be interpreted, alternatively, as a functional defined on
$\aa(\ep)$. However, we are interested less in the functional
(66) itself than in the fact that for a {\it given} 
Dirac operator $\dit\!\in\!\sdi$ the Wodzicki residue
of $\dit^{-2n+2}$ can be considered as a functional
\bb
{\cal I}_{\rm D}'(\!\nat\!) := {\rm res}(\dit^{-2n+2}),
\ee
of all connections $\nat\!\in\!\aa(\ep)$ so that 
$c(\!\nat\!)=\dit$. In other words: with respect to a given 
Dirac operator (68) can be considered as a certain functional
on the subspace of all (endomorphism valued) differential 
forms defining this Dirac operator. In the case that $\dit$ 
denotes a SDO the functional (68) is proportional to the 
Einstein-Hilbert action. We again stress that this was 
recognized by Connes, c.f. \cite{Co2} and was proved in 
\cite{K1}. From a more general point of view (see below) this 
was also discovered in \cite{KW}, which in turn was the 
starting point to deal with non-SDO's in \cite{AT1} and 
\cite{AT2}.

There is still another motivation for (68); the connection of 
(68) to the heat trace associated with (the square of)
a Dirac operator. For this let us remind that there 
is a natural functional on $\sdi$
\bb
\sdi &\rightarrow&\cc \cr
\dit &\mapsto& {\rm Tr}\,e^{-\tau\pdit^2}.
\ee
Though in general one is not able to calculate this 
functional, it is well-known that it has an asymptotic 
expansion:
\bb
{\rm Tr}\,e^{-\tau\pdit^2}\sim (4\pi\tau)^{-n}
\sum_{k\geq 0}\tau^{(k - 2n)/2}\sigma_k(\pdit^2),
\ee 
where the coefficients (Seeley-DeWitt coefficients) 
\bb
\sigma_k(\pdit^2) :=
\int_\mm\!\ast\hbox{tr}_\ep\,\sigma_k(x;\pdit^2)
\ee
are known to contain geometric information. In particular, the 
subleading term $\sigma_2(\pdit^2)$ is of the general form 
(c.f. \cite{Gl}):
\bb
{\sigma}_2(\dit^2) = 
\int_\mm\!\ast\hbox{tr}_\ep\left[\hbox{$\frac{1}{6}r_\mm$} 
- {\cal F}\right]
\ee
and thus is proportional to (66). Of course, this is by no 
means accidental. In general, for all 
$(2n\!-\!k)/2\!\notin\!\zz$ ($\mm$ smooth) one has
( c.f. \cite{Wo2})
\bb
{\sigma}_k(\dit^2) = 
\hbox{$\frac{\Gamma((2n\!-\!k)/2)}{2}$}\;
{\rm res}(\dit^{-2n +k}).
\ee

From this point of view the statement of our main theorem
(as given in the introduction) may be rephrased as 
follows: 

{\sl There exists a Hamiltonian (generalized Laplacian)} 
${\cal H}$ {\sl such that the subleading term in the 
asymptotic expansion of the corresponding heat trace 
associated with this Hamiltonian is proportional to the 
classical bosonic action of the standard model with gravity 
including. Moreover, this Hamiltonian has a 
square root,} ${\cal H}\!=\!\dit^2,$ {\sl which gives rise also 
to the fermionic action of the standard model.}\\

In what follows we assume that $\mm$ denotes a {\it spin} 
manifold\footnote{Indeed, it is widely believed by physicists 
that fermions are geometrically described by spinors.}.
Although this is not necessary, since we are only interested 
in local objects (densities), it simplifies notation. 
Consequently, the (total space of the) Clifford module bundle 
globally takes the form: $\ep\!=\!{\rm S}\ot{\rm E}$. To get 
in touch with gauge theory we assume that E denotes
an associated (hermitian) vector bundle: 
${\rm E}\!=\!{\cal P}\times_\rho V$, where
${\cal P}$ is a G-principal bundle over $\mm$ and 
$\rho\!:\!{\rm G}\rightarrow V$ is a unitary representation 
of the (real, compact and semi-simple) Lie-group G on a 
hermitian vector space $V$. By ${\cal G}$ we denote the 
corresponding Lie-algebra of G and by $\rho'$ the induced 
representation of ${\cal G}$ on ${\rm End}(V)$. Consequently, 
any connection form A on E takes its values in 
$\rho'({\cal G})\subset{\rm End}(V)$. This offers the 
possibility of defining the slightly more general 
functional\footnote{That this functional actually is more 
adequate than (68) becomes clear when discussing the 
physical implications of the proposed model, c.f. part 2.},
\bb
{\cal I}_{\rm D}(\!\naep\!) &:=& 
{\rm res}_\zeta(\dit^{-2n + 2})\nonumber\\[.5em]
&\equiv& {\rm res}\left(\,(\zeta\,\dit^2)^{-n+1}\right).
\ee
Here, $\zeta\!\in\!\Gamma(\enep)$ denotes an element
of the commutant defined by $({\rm G}, \rho)$. More precisely.
Let us recall the simple fact that any section 
$s\!\in\!\Gamma({\rm E})$ in an associated vector bundle
${\rm E}\!=\!{\cal P}\times_\rho V$ uniquely 
corresponds to an equivariant section 
${\bar s}\!\in\!\Gamma_{\!aq}({\cal P},{\rm V})$. 
Here, equivariant means: 
${\bar s}(pg)\! =\!\rho^{-1}(g){\bar s}(p),
\forall\, g\!\in\!{\rm G},\, p\!\in\!{\cal P}$. We then have the 
following\\ 

\noindent
{\bf Definition 7:} For a given generalized Dirac operator, 
consisting of the triple 
\bb
({\rm G}, \rho, \dit)
\ee 
with $\dit\!\in\!\sdi$, let us denote by 
${\bar z}\!\in\!\Gamma_{\!aq}({\cal P},{\rm End}({\rm V}))$ 
an element of the {\it commutant} 
\bb
{\cal C}_\rho({\rm G}) := 
\left\{a\in{\rm End}({\rm V})\,|\,[\rho(g), a] = 0, 
\forall\, g\!\in\!{\rm G}\,\right\}
\ee
and by $z\!\in\!\Gamma({\rm End}({\rm E}))$ its 
corresponding section in the endomorphism bundle 
associated with E\footnote{In the following we shall not 
distinguish between $s$ and ${\bar s}$.}. Since 
${\rm End}({\rm S})$ is simple, this generalizes
to $\enep$ via $\zeta\!: =\!{\bf 1}_{{\petit{\rm S}}}\!\ot\! z$. 
We impose the following three conditions on $\zeta$: it is a 
positive operator ($\zeta\!>\! 0$) and satisfies 
$[\dit, \zeta] = [\chi, \zeta] = 0$, where 
$\chi\in\Gamma({\rm End}(\ep))$ denotes the grading  
operator on $\ep\!=\!\ep^+\!\op\!\ep^-$.\\

\noindent
{\bf Lemma 5:} {\it As a consequence, $\zeta^\alpha$ has a 
constant spectrum (i.e. it is independent of $x\!\in\!\mm$) 
and the operator: $\zeta^\alpha\di^2$ is elliptic for any 
power $\alpha\!\in\!\rr$. Using this, we obtain} 
\bb
{\rm res}_\zeta(\dit^{-2n+2}) = 
\hbox{$\frac{2}{{\petit{\Gamma(n\!-\!1)}}}$}
\int_\mm\!\ast{\rm tr}_\ep
\left(\zeta\sigma_2(x; \pdit^2)\right).
\ee\\

\noindent
{\sl Proof:} Let us denote by
${\rm spec}\zeta(x)\! =\!
\{\lambda_1\cdots\lambda_k\}|_x,\; x\!\in\!\mm$ 
the spectrum of the positive operator 
$\zeta\!\in\!\Gamma({\rm End}(\ep))$. Here, $k$ indicates
the number of distinct eigenvalues of $\zeta$. We first 
prove that the spectrum is independent of $x\!\in\!\mm$.\\
\noindent
Since $z\!\in\!{\cal C}_\rho({\rm G})$ this section is
gauge invariant. I.e. for all gauge transformations
$f\!\in\!\Gamma_{eq}({\rm Aut}({\cal P}))\!\simeq\!
\Gamma({\cal P}\!\times_{ad}\!{\rm G})$ of ${\cal P}$ we
have: $f^* z\! =\!z$. Hence, it is sufficient to consider a
local situation. Using the fact that locally, any Dirac operator 
$\dit\!\in\!\sdi$ may be written as
\bb
\dit = c(\de) +\sum^m_{j=0}c({\cal A}_{[j]})
\ee
it follows that $[\dit,\zeta] = 0$ is equivalent 
to
\bb
\left[ c({\cal A}_{[j]}),\zeta\right] &=& 0, \quad
\forall j = 0,2,\cdots,m\, ,\nonumber\\[.5em]
c(\de\zeta) &=& -\left[c({\cal A}_{[1]}),\zeta\right].
\ee
However, the latter implies: $\de z\!=\!0$. Indeed, 
${\cal A}_{[1]}\! =\! \omega^S\!\ot\!{\bf 1}_{{\petit{\rm E}}}
\!+\!{\bf 1}_{{\petit{\rm S}}}\!\ot\!{\rm A}$, 
${\rm A}\!\in\!\Omega^1(\mm,\rho'({\cal G}))$ and thus 
$[{\rm A}, z]\!=\!0$. Consequently, $\zeta$ must be constant, 
which yields the first assertion.\\ 

\noindent
Since the spectrum of $z$ is constant we get\footnote{I like
to thank M. Lesch for explaining me this calculation.}
\bb
{\rm E} &=&\oplus^k_{j=1}
{\rm ker}(z -\lambda_j)\nonumber\\[.5em]
&=:&\oplus^k_{j=1}{\rm E}_j.
\ee
Hence,
\bb
\zeta &=& 
\oplus^k_{j=1}
\lambda_j{\bf 1}_{{\petit{\ep}}j}\,,\nonumber\\[.5em]
\di &=&\oplus^k_{j=1}\di_j,
\ee
where $\dit_j$ is the restriction of 
$\dit$ to $\ep_j\! :=\! {\rm S}\ot{\rm E}_j$. Consequently, we 
end up with
\bb
\zeta\dit^2 = \oplus^k_{j=1}\lambda_j\dit^2_j\,,
\ee
which implies
\bb
{\rm Tr}\,e^{-\tau\zeta{\petit{\dit^2}}} &=& 
\sum_{j=1}^k{\rm Tr}\,e^{-\tau\lambda_j{\petit{\dit^2_j}}}\cr
&\sim&
\sum_{l\ge 0}\sum_{j=1}^k\tau^{\frac{(l-m)}{2}}\,
\lambda_j^{\frac{(l-m)}{2}}\int_\mm\!\ast{\rm tr}_{\ep_j}\,
\sigma_l(x, \pdit_j^2),
\ee
and thus proves the lemma, when $\zeta$ is replaced by 
$\zeta^{-n+1}$.\\

\bigskip

Since we now have fixed the mathematical frame we conclude
this part by summarizing the proposed model building kit 
as follows: Let
\bb
({\rm G}, \rho, \dit)
\ee
be a given generalized Dirac operator defined on a Clifford
module bundle $\ep$. Then, the general action functional
on $\aa(\ep)\times\Gamma(\ep)$ is defined as
\bb
{\cal I}_{\tilde{\rm D}} &\equiv& 
{\cal I}_{fermionic}(\nat\!,\psi)\; +\; 
{\cal I}_{bosonic}(\!\nat\!)\nonumber\\[.5em]
&:=&
(\psi, \dit\psi)_{\Gamma(\ep)} +
{\rm res}_\zeta(\dit^{-2n + 2}),
\ee
where $c(\!\nat\!)=\dit$.\\

\section{Part 2: The Standard Model}

In this part of the paper we are concerned with the application
of the kit introduced in part 1 concerning the standard
model of particle physics. We therefore shall introduce in 
the following section an appropriate generalization of the 
Dirac-Yukawa operator and prove our main theorem.
Moreover, we shall discuss some consequences regarding
the various parameters involved in the model.\\ 

\subsection{The Dirac-Yukawa operator and the
EHYMH-Action}

To get started let us give the following\\

\noindent
{\bf Definition 8:} Let $\ep\!:=\!{\rm S}\ot{\rm E}$ be a 
Clifford module bundle with a twisting graduation 
(${\rm E}\!=\!{\rm E}_{\rm L}\op{\rm E}_{\rm R}$) and 
denote by\footnote{Later we shall be mostly interested in 
the case $n=2$.}
$\chi\!:=\!\gamma_5\!\ot\!\chi^{\rm E}$ the appropriate
grading operator: $\chi^2\!=\!{\bf 1}_\ep,\; \chi^*\!=\!\chi$.
A Dirac operator $\dipi$ is called a (euclidean) 
{\sl Dirac-Yukawa} operator if it takes the form
\bb
\dipi&:=&\di +
i\gamma_5\ot\pmatrix{ 0   & {\tilde\phi}\cr
                                 {\tilde\phi}^*  &  0  },\nonumber\\[.5em]
&\equiv&\di + \gamma_5\ot\phi
\ee
where $\di$ is a SDO and ${\tilde\phi}\!\in\!
\Gamma({\rm Hom}({\rm E}_{\rm R}, {\rm E}_{\rm L}))$.\\ 

Since the Yukawa coupling (9) geometrically can be 
considered as defining a particular section $\phi$
(see below) one may try to naturally 
generalize the operator (86) in such a way that it not only
defines the fermionic action (8-9) but also yields the 
bosonic action (10-11). Here, "naturally" means that the 
generalization of (86) is determined by those elements only
which already determine the Dirac-Yukawa operator, i.e. by 
$(g,\phi,{\rm A})$.\\

\noindent
{\bf Theorem 1:} {\it Let $\ep\!=\!{\rm S}\op{\rm E}$ be a 
twisted Clifford module over $\mm$, with 
${\rm E}\!:=\!{\rm E}_{\rm L}\op{\rm E}_{\rm R}$. The
functional (74) evaluated with respect to the Dirac operator
\bb
\dit_\phi:= \di + a_4\Phi + 
{\cal J}\left(a_2\,c({\rm F}^{\ep/{\rm S}}) 
+ a_3\,c(\nabla^\penep\Phi) + a_o\Phi^2\right)
\ee
defined on ${\tilde\ep}\!:=\!{\rm S}\!\ot\!{\tilde{\rm E}}$ 
yields
\bb
{\rm res}_\zeta\left(\dit^{-2n+2}_\phi\right) =
-\hbox{$\frac{{\rm tr}_\ep\zeta}{3\,\Gamma(n-1)}$}\,
\int_\mm\!\ast\Big\{\,r_\mm\! &+&\! 
a'_4\,{\rm tr}_{\rm E}(z\phi^2)\nonumber\\[.5em]
\!&-&\! a'_2\,
{\rm tr}_{\rm E}(z{\rm F}_{\!\mu\nu}
{\rm F}^{\mu\nu})\nonumber\\[.5em]
\!&+&\! a'_3\,
{\rm tr}_{\rm E}(z\nabla_{\!\mu}\phi
\nabla^\mu\phi)\nonumber\\[.5em]
\!&+&\! a'_o\,
{\rm tr}_{\rm E}(z\phi^4)\Big\},
\ee
with}
\bb
a'_o &:=& 
\frac{24n(1-\frac{1}{2n})}
{{\rm tr}_{\rm E}z}\,a_o^2,\nonumber\\[.5em]
a'_2 &:=& \frac{24(2n-3)}
{{\rm tr}_{\rm E}z}\,a_2^2,\nonumber\\[.5em]
a'_3 &:=& \frac{24(n-1)}
{{\rm tr}_{\rm E}z}\,a_3^2,\nonumber\\[.5em]
a'_4 &:=& \frac{12}{{\rm tr}_{\rm E}z}\,a_4^2.
\ee
Here, $\Phi\!:=\!\gamma_5\!\ot\!\phi$ and
${\rm F}\!\in\!\Omega^2(\mm,\rho'({\cal G}))$ is the
Yang-Mills curvature, induced by the gauge potential
${\rm A}$ in the definition of the SDO $\di\!\in\!
{{\cal D}(\ep)}$. The covariant derivative $\nabla\phi$ 
is defined with respect to the adjoint representation of G, 
where the $\phi$ sits in\footnote{I.e. the corresponding 
equivariant section 
${\bar\phi}\!\in\!\Gamma_{eq}({\cal P},{\rm End}(V))$
fulfils: ${\bar\phi}(pg)\!=\!\rho^*(g){\bar\phi}(p)\rho(g),\,
\forall\,p\!\in\!{\cal P},\,g\!\in\!{\rm G}.$}.
The a's denote arbitrary (complex) constants. The structure
group G is assumed to act on 
${\tilde{\rm E}}:={\rm E}\!\op\!{\rm E}$ via the 
representation ${\tilde\rho}:=\rho\!\op\!\rho$, and the 
automorphism 
${\tilde{\rm E}}\mapright{{\cal J}}{\tilde{\rm E}}$ denotes
the corresponding "complex structure" on ${\tilde{\rm E}}$. 
Moreover, we have used the canonical identification 
\bb
A + \,{\cal J}(B) = \hbox{$\pmatrix{A  &  -B\cr
                                           B  &  A}$}
\ee
for all $A, B\!\in\!{\rm End}({\rm E})$. Note, in what
follows we do not distinguish between the Clifford action 
$c$ on the Clifford module $\ep$ and the corresponding 
action ${\tilde c}:= c\!\op\!c$ on the (canonically) induced 
Clifford module ${\tilde\ep}$. Likewise, we do not 
distinguish between the representation $\rho$ and 
${\tilde\rho}$.\\

Consequently, if the constant $a_3$ is purely
imaginary and the other constants are real, the functional
(74) with respect to the generalized Dirac-Yukawa operator
(87), becomes proportional to the 
Einstein-Hilbert-Yang-Mills-Higgs action (EHYMH) of the 
standard model. Moreover, if one considers 
"diagonal sections": 
${\tilde\psi}\!:=\!(\psi,\psi), \psi\!\in\!\Gamma(\ep)$ only, 
the fermionic functional in (85) becomes proportional to the 
Dirac-Yukawa action. Note, in (85) there is still a length 
scale missing because the fields do not yet have the right 
dimensions. This will be discussed in the next section where 
we shall consider some constraints of our approach to the 
standard model.\\

\noindent
{\bf Remark 7:} The (euclidean) Dirac-Yukawa operator (86), 
is uniquely defined by the Clifford superconnection
\bb
{\nabla\!\!\!\!\nabla}^{\rm E}&:=&\nabla^{\rm E}+\epsilon,\cr
\epsilon&:=&i^n\ast\!\phi
\ee
on E, where we have used the fact that the grading operator 
$\gamma_5$ of the spinor bundle S is proportional to the 
volume form on $\mm$. More precisely, we have
$c(i^n\ast\!1)\!=\!
\gamma_5\ot{\bf 1}_{\rm E}\!\in\!
\Gamma({\rm End}^+(\ep))$.\\

Hence, the (euclidean) Dirac-Yukawa operator is determined 
by a 1-form (gauge potential) and the 2n-form in (91). In 
four dimensions, however, the most general Dirac operator 
in addition depends on a zero form, a two form and a three
form. By considering only those forms which already determine the 
Dirac-Yukawa operator (91) naturally yields the following ansatz
\bb
{\nabla\!\!\!\!\nabla}^{\tilde{\rm E}}&:=& \nabla^{\rm E} + 
a_4\,\epsilon\; + {\cal J}\left( a_2\,{\rm F} 
- a_3\,\ast\!\nabla^{{\rm End}({\rm E})}(\ast\epsilon) 
+ a_o\,(\ast\epsilon)^2\right)
\ee
on ${\tilde{\rm E}}$. Note that the forms within the brackets
are even with respect to the total degree and hence
${\cal J}(\cdots)$ becomes odd (c.f. also the final section.). 
It is easily checked that
\bb
\dit_{{\nabla\!\!\!\!\nabla}^{\tilde\ep}} = \dit_\phi
\ee
with ${\nabla\!\!\!\!\nabla}^{\tilde\ep}\!:=\!
{\nabla}^{\rm S}\!\ot\!{\bf 1}_{\rm E}\!+\!
{\bf 1}_{\rm S}\!\ot\!{\nabla\!\!\!\!\nabla}^{\tilde{\rm E}}.$\\

\noindent
There are two interesting choices for the constants $a_k$ so 
that (87) takes a particularly nice geometric
form: First, the most natural choice is
$a_k\!\equiv\!1, \forall\,k\!=\!0,2\cdots 4$. In this case  
the Dirac operator (87) reads
\bb
\dit_\phi &=& c(\naep + \Phi) + 
{\cal J}\left(c(
{\cal R}_{{\tilde{\nabla\!\!\!\!\nabla}}^{\ep}}
)\right)\nonumber\\[.5em]
&=:&c({\tilde{\nabla\!\!\!\!\nabla}}^{\ep}) +
{\cal J}\left(c(
{\cal R}_{{\tilde{\nabla\!\!\!\!\nabla}}^{\ep}}
)\right)\nonumber\\[.5em]
&=& \di_\phi +
{\cal J}\left(c(
{\cal R}_{{\tilde{\nabla\!\!\!\!\nabla}}^{\ep}}
)\right)\nonumber\\[.5em]
&\equiv&\di_{{\tilde{\nabla\!\!\!\!\nabla}^{\tilde\ep}}}
\ee
with the "super relative curvature" 
${\cal R}_{{\tilde{\nabla\!\!\!\!\nabla}}^{\ep}}\!:=
\!({\tilde{\nabla\!\!\!\!\nabla}}^{\ep})^2 - 
(\nabla^{\rm S})^2\!\ot\!{\bf 1}_{\rm E}$.\\ 
However, there is still another nice choice: 
$a_o\!=\!a_3\!:=\!(1\!-\!\frac{1}{2n}),\,
a_2\!=\!a_4\!:=\!1$; in this case the generalization (87)
of the Dirac-Yukawa operator (86) reads
\bb
\dit_\phi &=& c(\naep + \omega_\phi) +
{\cal J}\left(c({\rm F}^{\ep/{\rm S}} + 
\de^{\!\pnaep}\!\!\omega_\phi + 
\omega_\phi\wedge\omega_\phi)\right)\nonumber\\[.5em]
&=:& c(\nat) + 
{\cal J}\left(c({\cal R}_{{\tilde\nabla}^\ep}
)\right)\nonumber\\[.5em]
&=& \di_\phi + 
{\cal J}\left(c({\cal R}_{{\tilde\nabla}^{\ep}}
)\right)\nonumber\\[.5em]
&\equiv& \di_{{{\tilde\nabla}^{\tilde\ep}}},
\ee
with the "{\it Higgs-form}" 
$\omega_\phi\!:=\!\delta_\xi\Phi\!\in\!
\Omega^1(\mm,{\rm End}^+(\ep)\,)$ and
the relative curvature 
${\cal R}_{{\tilde\nabla}^{\ep}}\!:=\!(\!\nat\!)^2\! -\! 
(\nabla^{\rm S})^2\!\ot\!{\bf 1}_{\rm E}$. Because of
\bb
\di_{{\tilde{\nabla\!\!\!\!\nabla}}^{\tilde\ep}} -
\di_{{{\tilde\nabla}^{\tilde\ep}}} =
\hbox{$\frac{1}{2n}$}\,{\cal J}
\left(c({\cal R}_{{\tilde{\nabla\!\!\!\!\nabla}}^{\ep}} - 
{\rm F}^{\ep/{\rm S}})\right),
\ee
the  Dirac operators 
$\di_{{\tilde{\nabla\!\!\!\!\nabla}}^{\tilde\ep}},
\di_{{\tilde\nabla}^{\tilde\ep}}$, however, are different. 
Therefore, strictly speaking, the definition (87) gives a 
whole class of Dirac operators, parametrized by the 
constants $a_o,\ldots, a_4$, and which all yielding (88).\\

After this remark let us turn to the proof of the theorem.\\

\noindent
{\sl Proof:}
Clearly, to prove this theorem we just have to calculate the
endomorphism (47) with respect to the Dirac operator (87). 
This tedious but straightforward calculation can most easily 
be achieved using the local form of (47):
\bb
{\cal F}^{{\tilde\nabla}^{\tilde\ep}} &=&
\hbox{$\frac{1}{4}$}\,r_\mm{\bf 1}_{\tilde\ep} +
\hbox{$\frac{1}{2}$}\,
\gamma^{\mu\nu}\ot{\rm F}_{\!\mu\nu}\nonumber\\[.5em]
&+&\hbox{$\frac{1}{2}$}\,
[\gamma^\mu[\omega_\mu,\gamma^\nu],\omega_\nu] +
\gamma^{\mu\nu}('\nabla_{\!\mu}\omega_\nu) -
\hbox{$\frac{1}{2}$}\,
\gamma^\mu[('\nabla_{\!\nu}\omega_\mu),\gamma^\nu] 
\nonumber\\[.5em]
&+& \hbox{$\frac{1}{2}$}\,
\gamma^{\mu\nu}[\omega_\mu,\omega_\nu] +
\hbox{$\frac{1}{4}$}\,g_{\mu\nu}
\gamma^\alpha[\omega_\alpha,\gamma^\mu]
\gamma^\beta[\omega_\beta,\gamma^\nu],
\ee
where $\cb_\cc\mapright{\gamma}{\rm End}({\rm S})$ denotes
the (chiral) representation of the complexified Clifford 
algebra and $\gamma^\mu\!\equiv\!c\gamma(e^\mu)$ with
$\{e^\mu\}_{1\leq\mu\leq 2n}$ a local frame in $T^*\!\mm$.
Also we use the shorthand notation 
$\gamma^{\mu\nu}\!:=\![\gamma^\mu,\gamma^\nu]/2$.
Moreover, F is the curvature on E, associated to the 
connection form A. The connection 
${\tilde\nabla}^{\tilde\ep}\!\in\!\aa({\tilde\ep})$ denotes
any representative  of the corresponding class defining the
Dirac operator (87) and $\omega\!:=\!
{\tilde\nabla}^{\tilde\ep}\!-\!{\nabla^{\tilde\ep}},$ with
${\nabla^{\tilde\ep}}\!\in\!\aa_{Cl}({\tilde\ep})$ arbitrarily
chosen, and $'\nabla_{\!\mu}$ denotes the 
covariant derivative induced by this Clifford connection on
${\rm End}({\tilde\ep})$. 

In particular, we may choose ${\tilde\nabla}^{\tilde\ep}$ 
such that
\bb
\omega &:=& \omega^o +\omega^2 +\omega^3 +
\omega^4\quad\hbox{with}\nonumber\\[.5em]
\omega^o &:=& a_o\,\delta_\xi{\cal J}(\Phi^2),\cr
\omega^2 &:=& 
a_2\,e^\mu\ot\gamma^\nu\ot{\cal J}({\rm F}_{\mu\nu}), \cr
\omega^3 &:=& a_3\,{\cal J}(\nabla\Phi), \cr
\omega^4 &:=& a_4\,\delta_\xi(\Phi).
\ee
Note, we have already omitted  
$\omega^1\!:=\!\omega^{\rm S}\!\ot\!{\bf 1}_{{\rm E}} +
{\bf 1}_{\rm S}\!\ot\!{\rm A}$, which would change the
Clifford connection ${\nabla}^{\tilde\ep}$ only and not the 
functional (88)  by corollary 1.
The main advantage of the local form (97) is that the whole
calculation becomes purely algorithmic. Moreover, because
of the trace ${\rm tr}_{\tilde\ep}$ in (77) one only has to
calculate the last four terms in (97). For the same reason
most of the terms to be calculated will drop out. Indeed, 
using (98), it is easily checked that the two derivative terms 
in (97), actually, do not contribute to the functional (74). 
Hence, all it is left to be done is to calculate the last two quadratic 
terms in (97) up to "traceless" contributions. This 
calculation becomes even more simplified by the\\

\noindent
{\bf Remark 9:} Let $\ep_j,\,j\!=\!1,2$ be two Clifford modules 
over $\mm$ with the appropriate actions $c_j$. Then, 
$(\ep, c)$ with $\ep\!:=\!\ep_1\op\ep_1$ and 
$c\!:=\!c_1\op c_2$ is also a Clifford module. The most
general Dirac operator on this Clifford module takes the
form
\bb
\dit = \pmatrix{\dit_1       &    A_{12}\cr
                             A_{21}   &  \dit_2},
\ee 
where, respectively, 
$\dit_j\!\in\!{\cal D}({\ep_j}),\,j\!=\!1,2$,
$A_{12}\!\in\!\Gamma({\rm Hom}(\ep_2,\ep_1))$ and
$A_{21}\!\in\!\Gamma({\rm Hom}(\ep_1,\ep_2))$. The
Wodzicki residue of $\dit^{-2n+2}$ is such that there are
no terms "mixing the diagonal with the off-diagonal",
c.f. \cite{AT2}.\\

As a consequence, in calculating the corresponding quadradic 
terms of (97), products of the generic form 
$\omega^4\omega^k,\,k\!=\!0,\cdots, 3$ may be omitted as 
well. As a result we end up with
\bb
\hbox{$\frac{1}{2}$}\gamma^{\mu\nu}[\omega_\mu,\omega_\nu] &=&
a_o^2\,\hbox{$(1-\frac{1}{2n})$}\,\Phi^4 \nonumber\\[.5em]
&+& a_2^2\,
{\rm ev}_{\!g}\!\left({{\rm F}^{\ep/{\rm S}}}^2\right)
\nonumber\\[.5em]
&+&
a_4^2\,\hbox{$(1-\frac{1}{2n})$}\,\Phi^2
\ee
and
\bb
\hbox{$\frac{1}{4}$}\,g_{\mu\nu}
\gamma^\alpha[\omega_\alpha,\gamma^\mu]
\gamma^\beta[\omega_\beta,\gamma^\nu] &=&
2na_o^2\,\hbox{$(1-\frac{1}{2n})^2$}
\,\Phi^4\nonumber\\[.5em]
&+& 
(5-4n)\,a_2^2\,
{\rm ev}_{\!g}\!\left({{\rm F}^{\ep/{\rm S}}}^2\right)
\nonumber\\[.5em]
&+&
2(n-1)\,a_3^2\,{\rm ev}_{\!g}((\nabla^{\penep}\Phi)^2\,)
\nonumber\\[.5em]
&+&
\frac{1}{2n}\,a_4^2\,\Phi^2.
\ee
Note that both equalities hold up only to traceless 
terms! Finally, if we put all together, the  theorem is proven.\\

To summarize: we have introduced a certain Dirac operator 
(87), generalizing the Dirac-Yukawa operator (86), such that 
the functional (74) looks like that of the bosonic action of
the standard model with gravity including. This result is 
independent of the specific structure group G and its 
appropriate (fermionic) representation. To get in touch 
with physics, however, we also have to specify the pair 
$({\rm G}, \rho)$. In other words, we have to define the 
generalized Dirac operator of the standard model.\\

\subsection{The generalized Dirac-Yukawa operator of the 
standard model}
To begin with, we give the following\\

\noindent
{\bf Definition 9:}
Let $\ep$ be a twisted spinor bundle, so that
${\rm E} = {\rm E}_{\rm L}\!\op\!{\rm E}_{\rm R}$. We call 
\bb
({\rm G}, \rho, \di)
\ee
the {\sl generalized Dirac-Yukawa operator of the standard 
model} provided the structure group ${\rm G}$ takes the form
\bb
{\rm G} :=
{\rm SU}(3) \times {\rm SU}(2) \times {\rm U}(1)
\ee
and has the (fermionic) representation
$\rho\!:=\!\rho_{\rm L}\op\rho_{\rm R}\!: 
{\rm G}\rightarrow{\rm Aut}(V)$, 
\bb
\rho_{\rm L}({\bf c}, {\bf w},b) &:=&
\pmatrix{{\bf c}\ot{\bf 1}_{\rm N}\ot 
{\bf w}\,b^q_{\rm L}  &  0\cr
                            \phantom{a}  &  \phantom{b}\cr
0  & {\bf 1}_{\rm N}\ot {\bf w}\,b^l_{\rm L}},\\[.5cm]
\rho_{\rm R}({\bf c}, {\bf w},b) &:=&
\pmatrix{{\bf c}\ot{\bf 1}_{\rm N}\ot 
{\bf B}^q_{\rm R}  &  0\cr
                              \phantom{a}  &  \phantom{b}\cr
                                    0  &  {\bf B}^l_{\rm R}},
\ee
on the typical fiber
\bb
V &:=& V_{\rm L}\op V_{\rm R}\\[.2cm] 
&\simeq&
\left[(\cc^{6\rm N}_q\op\cc^{2\rm N}_l)\right]_{\rm L}\op
\left[(\cc^{3\rm N}\op\cc^{3\rm N})\!_q\op
\cc^{\rm N}_l\right]_{\rm R}\\[.2cm]
&\simeq&\cc^{8\rm N}_{\rm L} \op \cc^{7\rm N}_{\rm R}.
\ee
Moreover, the Dirac operator 
\bb
\di := \dit_\phi
\ee
is the generalized Dirac-Yukawa operator (87), 
where the homomorphism ${\tilde\phi}\!\in\!
\Gamma({\rm Hom}({\rm E}_{\rm R},{\rm E}_{\rm L}))$ 
is defined by
\bb
{\tilde\phi} &:=&
\pmatrix{{\bf 1}_3\ot({\bf g}'^q_y\ot\varphi,\, 
{\bf g}^q_y\ot\hbox{\boldmath$\epsilon$\unboldmath}
                    \overline{\varphi} ) & 0\cr
                    \phantom{a}  &  \phantom{b}\cr
                            0  &  {{\bf g}^l_y\ot\varphi}}\\[.5cm]
&\equiv&\pmatrix{{\bf 1}_3\ot{\tilde\varphi}_q & 0\cr                   
                                     0  &  {\tilde\varphi}_l}.
\ee
Here, respectively, 
${\bf g}'^q_y, {\bf g}^q_y\!\in\!{\bf M}_{\rm N}(\cc)$ denote
the matrices of the Yukawa coupling constants for quarks
of electrical charge -1/3 and 2/3 (i.e. of quarks of "d"-type, 
and of "u"-type) and ${\bf g}^l_y\in{\bf M}_{\rm N}(\cc)$
is the matrix of
the Yukawa coupling constants for the leptons of charge -1
(i.e. of leptons of "electron" type). While ${\bf g}^q_y$ 
and ${\bf g}^l_y$ can be assumed to be diagonal and real, 
the matrix ${\bf g}'^q_y$ is related to the Kobayashi-Maskawa 
matrix and therefore is neither diagonal nor real. The 
"weak hypercharges" for the left and right handed quarks 
(indicated by the subscript "q") and leptons (subscipt "l") 
are defined by: $\rho(b):=e^{iy\theta},\, b\in{\rm U}(1),\,
y\in\qqq,\, \theta\in[0,2\pi[$ (c.f. the introduction). Then the 
two by two diagonal matrices ${\bf B}^q_{\rm R}$ and 
${\bf B}^l_{\rm R}$ in the definition (105) are: 
${\bf B}^q_{\rm R}:={\rm diag}(b_{\rm R}^{d'}, b_{\rm R}^{u})$
and ${\,\bf B}^l_{\rm R}:=b^l_{\rm R}{\bf 1}_{\rm N}$.
In (110), $\varphi$ denotes a section into a rank two 
subbundle ${\rm E}_h$ of the vector bundle E and carries the 
defining representation $\rho_h$ of the electroweak subgroup
${\rm G}_h\!:=\!{\rm SU}(2)\!\times\!{\rm U}(1)$ of the 
structur group G. I.e., when $\varphi$ is considered as an 
element of $\Gamma_{\rm eq}({\cal P}, {\rm V})$ it 
transforms like $\varphi(pg) = \rho_h(g)^*\varphi(p)$
with  $g\!:=\!({\bf w}, b)\in{\rm G}_h\!=
\!{\rm SU}(2)\times {\rm U}(1)$ 
and $\rho_h({\bf w}, b):= {\bf w}\,e^{iy_h\theta}$. 
Finally, $\hbox{\boldmath$\epsilon$\unboldmath}$ is the 
anti-diagonal matrix 
$\hbox{\boldmath$\epsilon$\unboldmath}
:={\phantom{-}0\;1\choose-1\;0}$ and 
$\overline{\varphi}$ here means the complex conjugate of 
$\varphi$. \\ 
 
Hence, the Higgs field $\varphi$ defines an element
$\Phi$ in $\Omega^0({\mm, {\rm End}^-(\ep)})$ and thus
a Dirac-Yukawa operator (86). Of course, the particular form
of $\tilde\phi$ in (110) is such that $\Phi$ gives the correct 
Yukawa coupling term (9) in the definition of the fermionic
action of the standard model. However, since the section
$\phi\in\Gamma({\rm End}^-({\rm E}))$ transforms with 
respect to the ($\rho$-induced) adjoint representation of G it 
follows that the hypercharge $y_h$ of the Higgs field  and, 
respectively, the hypercharges $y^l_{\rm L}, y^l_{\rm R}$ of 
the left and right handed leptons and the hypercharges 
$y^q_{\rm L}, y^{d'}_{\rm R}, y^u_{\rm R}$ of the left and 
right handed quarks must satisfy the following well-known 
relations  
\bb
y_h &=& y^l_{\rm L} - y^l_{\rm R}\cr
                 &=& y^q_{\rm L} - y^{d'}_{\rm R}\cr
                 &=& y^u_{\rm R} - y^q_{\rm L}.
\ee
In other words, when the Yukawa coupling is of the form as 
defined by (9) the corresponding hypercharges are not all
independent but have to satisfy the relations (112).
Moreover, since we know the electrical charge of the
particles, the numerical values of the "$y'\hbox{s}$" are 
fixed: 
\bb
(y^q_{\rm L}, \,y^l_{\rm L}) &=& (1/6, -1/2),\\[.1cm]
(\,(y^{d'}_{\rm R}, y^u_{\rm R}),\,y^l_{\rm R}) &=&
(\,(-1/3,2/3),\,-1).
\ee
This is a consequence of the generalized Gell-Mann-Nishijima
relation
\bb
{\bf Q} = {\bf T}_3 + {\bf Y},
\ee
where $\{i{\bf T}_3\!:=
\!\rho'(\,(i\,\hbox{\boldmath$\tau$\unboldmath}_3)/2),\,  
i{\bf Y}\!:=\!\rho'(i)\}$ is a basis of the maximal Cartan
subalgebra of $\rho'({\rm su}(2)\op{u}(1)\,)$, so that 
$i{\bf Q}$ generates the residual structure group - in 
more physical terms the "electromagnetic gauge 
group"\footnote{As usual we use 
$\{{\bf E}_k\}_{1\leq k\leq 12}:=
\big\{(i\,\hbox{\boldmath$\lambda$\unboldmath}_a)/2,\,
(i\,\hbox{\boldmath$\tau$\unboldmath}_b)/2,\,
i\big\}_{1\leq a\leq 8\atop 1\leq b\leq 3}$
as a basis of ${\rm su}(3)\!\op\!{\rm su}(2)\!\op\!{\rm u}(1)$,
where 
$\hbox{\boldmath$\lambda$\unboldmath}_a,\, a\!=\!1\dots 8$
denote the Gell-Mann matrices and 
$\hbox{\boldmath$\tau$\unboldmath}_b,\, b\!=\!1\dots 3$ 
the Pauli matrices.}
\bb
{\rm U}_{\rm elm}(1)\subset{\rm SU}(2)\times{\rm U}(1)
\ee
in the fermionic representation $\rho$ after the mechanism of 
{\it spontanous symmetry breaking} is established. To make 
the latter more precise mathematically, let us remember how 
the notion of spontaneous symmetry breaking can be 
geometrically rephrased in terms of the reduction of a 
G-principal bundle, c.f. \cite{KN}, \cite{CBM}.\\

Let therefore ${\rm H}\!\subset\!{\rm G}$ be a Lie-subgroup 
of G and, respectively, ${\cal P}_{\rm G}$ and 
${\cal P}_{\rm H}$ be the (total spaces of the) corresponding 
principal bundles over the same base manifold $\cal M$. 
Then, ${\cal P}_{\rm H}$ is called an H-reduction of 
${\cal P}_{\rm G}$ iff 
${\cal P}_{\rm H}\!\subset\!{\cal P}_{\rm G}$ is a 
submanifold, so that the injection
${\cal P}_{\rm H}\hookrightarrow{\cal P}_{\rm G}$
is a bundle homomorphism. A necessary and sufficient
condition for a G-principal bundle ${\cal P}$ to be
H-reducible is that the ${\cal P}-$associated  fiber bundle 
${\cal P}_{\rm G}/{\rm H}\rightarrow{\cal M},$ with typical
fiber G/H, admits a global section.\\

\noindent
{\bf Definition 10:} A solution 
$(g, A, \varphi, \Psi)$ of the Euler-Lagrange equations 
corresponding to the functional 
(12) is called a {\sl classical vacuum} iff $g$ is
flat, $A =\Psi = 0$ and the Higgs field $\varphi = \varphi_o$ 
minimizes the Higgs potential 
\bb
&V:\Gamma({\rm E}_h)\longrightarrow\rr,&\cr
&V'|_{\varphi=\varphi_o} = 0.&
\ee
Here, ${\rm E}_h$ denotes the (total space of the) bundle
where the Higgs field lives in (see below) and, in principle, 
$V$ may be any gauge invariant polynomial of order less or 
equal then four of the Higgs field $\varphi$. Again, it is 
assumed that $V$ is bounded from below. Of course, 
concerning the standard model, $V$ has its well-known 
fashion (6).\\

Then, let us denote by $\Sigma\!\subset\!{\rm V}_h$ the 
(disjoint union of) orbits of classical vacuums with respect 
to the representation $\rho_h$ that carries the Higgs 
field $\varphi$ of the structure group ${\rm G}_h$.
In the special case of the standard model this group is 
identified with the "electroweak gauge group" 
${\rm SU}(2)\!\times\!{\rm U}(1)\!\subset{\rm G}$ and,
correspondingly, 
${\rm E}_h\!:=\!
{\cal P}\!\times_{\rho_h}\!{\rm V}_h\subset{\rm E}$
denotes the rank two subbundle describing, geometrically, 
the Higgs sector of the standard model. In the so-called 
"minimal version" of the standard model the typical fiber is
${\rm V}_h\simeq\cc^2$. Note that in the usual 
(non-geometrial) description of the standard model the
fermion representation $\rho$, defined by (104-108), and 
the Higgs representation $\rho_h$ are completely 
independent. Let us denote by 
${\rm I}(\varphi_o)\subset\rho_h({\rm G}_h)
\subset{\rm End}({\rm V}_h)$ the
isotropy group associated with a choosen classical vacuum
$\varphi_o\in\Sigma'\subset\Sigma,\,\hbox{connected}$.
Up to conjungation this isotropy group can be identified with
some Lie-subgroup H of ${\rm G}_h$ and we have
$\Sigma'\simeq(\,\rho_h({\rm G}_h)/\rho_h({\rm H})\,).$
Therefore, $\varphi_o$ considered as an element in 
$\Gamma_{eq}({\cal P}_h, {\rm V}_h)$ uniquely induces 
a section (also denoted by $\varphi_o$)  
${\cal M}\mapright{\varphi_o}{\cal P}_h/{\rm H}$. In other
words, from a geometrical point of view a necessary condition 
for a (classical) vaccum to exist is that the 
${\rm G}_h-$bundle must be H-reducible, where the 
Lie-subgroup ${\rm H}\subset{\rm G}_h$ is identified with
the isotropy group of some choosen Higgs field 
$\varphi_o$, minimizing the Higgs potential $V$.\\

\noindent
{\bf Definition 11:} The gauge symmetry is called 
{\sl spontaneously} broken by the (classical) vacuum,
represented by $\varphi_o$, iff 
${\rm H}\!\subset\!{\rm G}_h$ is a proper subgroup.\\

Note that the H-reduction of ${\cal P}_h$ is only necessary
but not sufficient for $\varphi_o$ to represent a classical
vacuum. It is also required that ${\cal P}_h$ posseses a flat
connection. In the particular case of the (minimal) standard
model we obtain:
$\Sigma = \{\rho_h({\bf w}, b)\varphi_o\}\,\coprod\,\{0\}$,
with $\varphi_o:=(0,\,v/\sqrt{2})^T$, 
$v\!:=\!\sqrt{\mu^2/\lambda}\in\rr$ and 
$({\bf w}, b)\in{\rm G}_h={\rm SU}(2)\!\times\!{\rm U}(1)$. 
Then, the isotropy group associated with the non-trivial 
$\varphi_o$ is generated by the anti-hermitian operator 
$i{\bf Q}$, and the corresponding residual structure
group H can be identified with ${\rm U}_{elm}(1)$, 
c.f. (115-116). Actually, the full residual structure 
group ("little group") of the standard model reads 
\bb
{\rm H} = {\rm SU}(3)\times{\rm U}_{elm}(1),
\ee
since the structure group G is given by (103).

Consequently, assuming that 
${\cal P}\!\equiv\!{\cal P}_{\rm G}$ 
is H-reducible and $\varphi_o$ represents an appropriate 
(non-trivial) classical vacuum of the standard 
model the section 
${\cal D}\!\in\!\Omega^0(\mm, {\rm End}^-(\rm E))$, defined
by
\bb
{\cal D}&\equiv&\phi_o\cr
            &:=& \phi|_{\varphi=\varphi_o}\nonumber\\[.5em]
            &=:& i\pmatrix{       0       & {\bf M}\cr
                                               {\bf M}^* & 0},
\ee
is H-invariant and thus well-defined on the reduced bundle.
Fixing the gauge so that $\varphi_o\!=\!(0,\!v/\sqrt{2})^T$
("unitary gauge") we may further write
\bb
{\bf M} &\equiv& \pmatrix{{\bf 1}_3\ot{\bf M}_q & 0\cr
     0 & {\bf M}_l},\;\hbox{with}\nonumber\\[.5em]
{\bf M}_q &=&\pmatrix{      0         & {\bf m}^{d'}\cr
{\bf m}^u  &         0},\\[.5em]
{\bf M}_l &=& {0 \choose {\bf m}^l}, 
\ee 
where, respectively, the matrices 
${\bf m}^l\!:=\!\frac{v}{\sqrt{2}}\,{\bf g}^l_y\!\in
\!{\bf{\rm M}}_{\rm N}(\cc)$ and 
${\bf m}^u\!:=\!\frac{v}{\sqrt{2}}\,{\bf g}^q_y\!\in
\!{\bf{\rm M}}_{\rm N}(\cc)$
denote the "mass matrices" of the charged leptons (l) and
quarks (q) of "u-type". They can be assumed
to be diagonal and real. The corresponding 
${\rm N}\!\times\!{\rm N}$ matrix 
${\bf m}^{d'}\!:=\!\frac{v}{\sqrt{2}}\,{\bf g}'^q_y$ of "d-type"
quarks is neither diagonal nor real. It is related to the
mass matrix of "d-type" quarks 
${\bf m}^d=\hbox{diag}(m^{d_1},\dots, m^{d_{\rm N}}),\,
m^{d_k}\!\in\!\rr,\,k\!=\!1,\dots, {\rm N}$ via the
Kobayashi-Maskawa matrix ${\bf V}\!\in\!{\rm U}(\rm N)$:
${\bf m}^{d'}\!=\!{\bf V}\,{\bf m}^{d}\,{\bf V}^*$.

Obviously, ${\bf M}$ denotes the {\it fermionic mass matrix} 
and we have recovered the "internal Dirac operator" ${\cal D}$ 
of the Connes-Lott approach to the standard model, c.f. 
\cite{CL} and the corresponding references therein. Again, we 
stress that ${\cal D}$ minimizes the Higgs potential $V$ but 
represents a classical vacuum only if ${\cal P}_{\rm H}$ 
(and thus ${\cal P}$) posesses a flat connection, c.f. 
\cite{CBM}.\\

As we have already mentioned, in the usual approach to the
standard model the representation $\rho_h$ of the Higgs field
$\varphi$ is independent of the fermionic representation 
$\rho_f$ as defined by (104-108). However, to be 
consistent one has to impose the relations (112) to the 
hypercharges. That these relations are not accidentally and, 
actually, must not be chosen by hand follows from the 
Dirac-Yukawa operator of the standard model: 
\bb
({\rm G},\,\rho,\,\dipi),
\ee
with $({\rm G},\,\rho\!\equiv\rho_f)$ like in (103-108) 
and $\dipi$ defined by (110). In other words: whenever one 
starts with (122) the representation of the Higgs field must 
be contained in the fermionic representation and then the 
relations between the hypercharge of the Higgs field and 
those of the corresponding fermions are fixed. If this does 
not hold, the Yukawa-coupling (9) would not define a 
generalized Dirac-operator.

We now turn to the consequences as implied by the 
generalization $({\rm G},\,\rho,\dit_\phi)$ of (122).
We therefore compare the functional (88), derived from 
the generalized Dirac-Yukawa operator (87), with the 
corresponding bosonic action of the standard model.
Before we can do this, however, we still have to give the 
various fields involved in the model their right dimensions, 
i.e., we first have to introduce an arbitrary length scale "$l$".
Also, we may choose to introduce a second endomorphism 
$\phi'$ on E in order to define the "off-diagonal" of (87). 
This additional endomorphism is defined by (110), but with 
the Yukawa-coupling matrices, 
${\bf g}'^q_y, \,{\bf g}^q_y,\,{\bf g}^l_y$ replaced by
arbitrary matrices
${\boldmath\hbox{$\Lambda$}\unboldmath}\!'^q, \,
{\boldmath\hbox{$\Lambda$}\unboldmath}\!^q,\,
{\boldmath\hbox{$\Lambda$}\unboldmath}\!^l$ 
of the corresponding size. This freedom arises from the
fact that the form of the functional (88) does not change by
this replacement and that the off-diagonal of $\dit_\phi$  
does not act on the fermions $\Psi$, giving the matrices 
${\bf g}_y$ their physical interpretation\footnote{At least, 
when one restricts oneself to diagonal sections: 
${\tilde\Psi} =(\Psi, \Psi)\in\Gamma({\tilde\ep})$.
The more general case of non-diagonal sections will be
discussed in a future paper, when also further 
phenomenological consequences of our model and a 
physical interpretation of the doubling of the "internal 
freedoms": 
${\rm E}\rightarrow{\tilde{\rm E}}\!=\!{\rm E}\!\op\!{\rm E}$ 
are considered, c.f. our remarks at the end of this 
paper.}. Note that this freedom is crucial for the definition 
of the mass matrix of the gauge bosons, see below. 
Moreover, without loss of generality we may assume that 
$\phi'$ is hermitian, so that all the constants $"a"$ are 
real, c.f. (89). With these replacements the "bosonic part"
of the universal functional (85) reads
\bb
{\cal I}_{bosonic} = {\cal I}_{EH} 
&+&\alpha_2\,\int_\mm{\rm tr}_{\rm E}
({\bf {\rm z}}\,F\wedge\ast F)\\[.5em]
&+&\alpha_3\,\int_\mm{\rm tr}_{\rm E}
({\bf{\rm z}}\,(\nabla\phi')^*\wedge
\ast(\nabla\phi')\,)\\[.5em]
&-&\alpha_4\,\int_\mm\ast{\rm tr}_{\rm E}
({\bf {\rm z}}\,\phi^*\phi)\\[.5em]
&+&\alpha_o\,\int_\mm\ast{\rm tr}_{\rm E}
({\bf {\rm z}}\,(\phi'^*\phi')^2\,)
\ee
with the constants
\bb
\alpha_o &:=& 
\frac{3n(1-\frac{1}{2n})}{2\pi\,{\rm tr}_{\rm E}{\bf{\rm z}}}
\,\left(\frac{l}{l_p}\right)^2 a_o^2,\\[.5em]
\alpha_2 &:=& 
\frac{3(2n-3)}{\pi\,{\rm tr}_{\rm E}{\bf{\rm z}}}
\,\left(\frac{l}{l_p}\right)^2 a_2^2,\\[.5em]
\alpha_3 &:=& 
\frac{3(n-1)}{2\pi\,{\rm tr}_{\rm E}{\bf{\rm z}}}
\,\left(\frac{l}{l_p}\right)^2 a_3^2,\\[.5em]
\alpha_4 &:=& 
\frac{3}{4\pi\,{\rm tr}_{\rm E}{\bf{\rm z}}}
\; m_p^2\, a_4^2.
\ee
Here, ${\bf{\rm z}}$ is considered as an element of
$\Gamma_{eq}({\cal P}, {\rm End}({\rm V}))$ that lies in 
the commutant (76) and satisfies: 
\bb
[{\bf{\rm z}},\chi^{\petit{\rm E}}] = [{\bf{\rm z}},\phi] =
[{\bf{\rm z}},\phi'] = 0,
\ee
as well as ${\bf{\rm z}}>0$.
 
Since we use our units so that $c\! =\!\hbar\!=\!1$ we have 
identified Newtons gravitational constant $G$ with the 
(square of the) "Planck length" $l_p\!\equiv\!m_p^{-1}$. 
Moreover, we already have normalized the Einstein-Hilbert 
functional ${\cal I}_{EH}$ so that 
\bb
{\cal I}_{EH} = \hbox{$\frac{1}{16\pi\,l_p^2}$}\,
\int_\mm\ast r_\mm.
\ee
The corresponding normalized fermionic action 
reads\footnote{Again, when restricted to diagonal sections.}
\bb
{\cal I}_{fermionic} &=& 
\int_\mm\ast(\Psi, i\di\Psi)_\ep\\[.5em]
& & a_4\,\int_\mm\ast(\Psi, i\Phi\Psi)_\ep,
\ee
with $\Phi$ defined by (110) and $\di$ a SDO. Note that 
only the constant $\alpha_4$ carries a dimension.

We now can compare the derived functionals 
(123-126) and (133-134) with the corresponding 
bosonic and fermionic action functionals of the standard 
model 
\bb
{\cal I}_{EHYMH} &:=& 
\hbox{$\frac{1}{16\pi l_p^2}$}
\int_\mm\!\ast r_\mm \\[.5em]
&+&
\int_\mm\!\ast
\left\{\hbox{$\frac{1}{2}$}{\rm tr}
({\bf C}_{\!\mu\nu}{\bf C}^{\!\mu\nu})
+ \hbox{$\frac{1}{2}$}{\rm tr}
({\bf W}_{\!\mu\nu}{\bf W}^{\!\mu\nu})
+ \hbox{$\frac{1}{4}$}
B_{\!\mu\nu}B^{\!\mu\nu}\right\} \\[.5em]
&+&
\int_\mm\!
\ast(\nabla\!\!_{\mu}\varphi)^*(\nabla^{\mu}\varphi) \\[.5em]
&+&
\int_\mm\!\ast
\left[\lambda(\varphi^*\varphi)^2 -
\mu^2\varphi^*\varphi\right]
\ee
and
\bb
{\cal I}_{DY} &=&\int_\mm
\ast(\Psi, i\gamma^\mu\nabla\!\!_{\mu}\Psi)_\ep\\[.5em]
&+& \,\int_\mm\ast(\Psi, i\Phi\Psi)_\ep.
\ee
Here, respectively,
\bb
{\bf C}_{\!\mu\nu}& :=& \partial\!_\mu {\bf C}_\nu - 
\partial\!_\nu {\bf C}_\mu
+ ig_{(3)}\,[{\bf C}_\mu, {\bf C}_\nu], \cr
{\bf W}_{\!\!\mu\nu}& :=& \partial\!_\mu {\bf W}\!\!_\nu - 
\partial\!_\nu {\bf W}\!\!_\mu
+ ig_{(2)}\,[{\bf W}\!\!_\mu, {\bf W}\!\!_\nu], \cr
B_{\mu\nu} &: =& \partial\!_\mu B_\nu - 
\partial\!_\nu B_\mu 
\ee
denote the ${\rm su}(3), {\rm su}(2)$ and 
${\rm u}(1)$ valued curvatures with respect to a local 
coordinate system and in the fundamental representation. 
Moreover, the covariant derivatives, acting on the Higgs
field $\varphi$ and on the fermions $\Psi$, are locally 
defined as
\bb
\nabla\!_\mu\,\varphi &:=&
(\partial\!_\mu  + ig_{(2)}\, W^b\!\!_\mu\,
\frac{\hbox{\boldmath$\tau$\unboldmath}_b}{2} 
+ ig_{(1)}y_h\,B_\mu)\varphi,\\[.3em]
\nabla\!_\mu\,\Psi &:=&
(\partial\!_\mu  +
ig_{(3)}\,C^a\!\!\!_\mu\;{\bf 1}_{\rm S}\ot{\bf F}\!_a +
ig_{(2)}\,W^b\!\!_\mu\;{\bf 1}_{\rm S}\ot{\bf T}_b  + 
ig_{(1)}\,B_\mu\;{\bf 1}_{\rm S}\ot{\bf Y})\Psi,
\ee
where again $(g_{(3)}, g_{(2)}, g_{(1)})$ are the 
coupling constants and $\left\{i{\bf F}\!_a, i{\bf T}_b, 
i{\bf Y}\right\}_{1\leq a\leq 8\atop 1\leq b\leq 3}$ denote
the generators of 
$\rho({\rm SU}(3)\times{\rm SU}(2)\times{\rm U}(1))$; 
$\lambda, \mu^2\,>0$ are the positive real constants, 
parametrizing the classical vacuum. Note that the curvatures 
(141) are hermitian.\\ 

\noindent
{\bf Lemma 6:} {\it The derived functionals (123-126) and 
(133-134) are identical with the bosonic and fermionic 
action functionals of the
standard model (135-140), iff the following relations 
hold:
\bb
{\rm N}\alpha_2\,g_{(1)}^2 &=&
\frac{1}{2(3y_q\lambda_q + y_l\lambda_l)},\\[.5em]
{\rm N}\alpha_2\,g_{(2)}^2 &=&
\frac{1}{(3\lambda_q + \lambda_l)},\\[.5em]
{\rm N}\alpha_2\,g_{(3)}^2 &=&
\frac{1}{4\lambda_q},
\ee
and
\bb
2\alpha_o\left[3\lambda_q\,
{\rm tr}({\hbox{\boldmath$\Lambda$\unboldmath}^*_q}
{\hbox{\boldmath$\Lambda$\unboldmath}_q})^2 +
\lambda_l\,
{\rm tr}({\hbox{\boldmath$\Lambda$\unboldmath}^*_l}
{\hbox{\boldmath$\Lambda$\unboldmath}_l})^2\,\right] 
&=& \lambda,\\[.5em]
2\alpha_3\left[3\lambda_q\,
{\rm tr}{\hbox{\boldmath$\Lambda$\unboldmath}^*_q}
{\hbox{\boldmath$\Lambda$\unboldmath}_q} +
\lambda_l\,
{\rm tr}{\hbox{\boldmath$\Lambda$\unboldmath}^*_l}
{\hbox{\boldmath$\Lambda$\unboldmath}_l}\,\right]
 &=& 1,\\[.5em]
2\alpha_4\left[3\lambda_q\,
{\rm tr}{\bf g}^*_q{\bf g}_q +
\lambda_l\,
{\rm tr}{\bf g}^*_l{\bf g}_l\right] &=& \mu^2,\\[.5em]
a_4 &=& 1
\ee
with the abbreviations}
\bb
y_q &:=& 2(y^{q}_{\rm L})^2 + (y^{d'}_{\rm R})^2 +
(y^{d'}_{\rm R})^2\in\rr_+,\nonumber\\[.5em]
y_l &:=& 2(y^{l}_{\rm L})^2 + (y^{l}_{\rm R})^2\in\rr_+,
\nonumber\\[.5em] 
{\hbox{\boldmath$\Lambda$\unboldmath}}_q &:=&
\left({\hbox{\boldmath$\Lambda$\unboldmath}}'^{q},
 {\hbox{\boldmath$\Lambda$\unboldmath}}^{q}\right)
\in{\bf M}_{2{\rm N}\!\times\!{\rm N}}(\cc),
\nonumber\\[.5em]
{\bf g}_q &:=&
\left({\bf g}'^{q}, {\bf g}^{q}\right)
\in{\bf M}_{2{\rm N}\!\times\!{\rm N}}(\cc),
\nonumber\\[.5em]
{\hbox{\boldmath$\Lambda$\unboldmath}_l} &:=&
{\hbox{\boldmath$\Lambda$\unboldmath}}^{l}
\in{\bf M}_{{\rm N}}(\cc),\nonumber\\[.5em]
{\bf g}_l &:=&{\bf g}^{l}\in{\bf M}_{{\rm N}}(\rr).
\ee\\
\noindent
{\sl Proof:} The proof of this lemma mainly consists in 
determining ${\bf{\rm z}}$ of the commutant, yielding 
\bb
{\bf{\rm z}} &=&\pmatrix{{\bf{\rm z}}_{\rm L} & 0\cr
0 & {\bf{\rm z}}_{\rm R}},\\[.5em]
{\bf{\rm z}}_{\rm L}&:=&
\pmatrix{\lambda_q\,{\bf 1}_{6\rm N} & 0\cr
0 & \lambda_l\,{\bf 1}_{2\rm N}}\\[.3em]
{\bf{\rm z}}_{\rm R}&:=&
\pmatrix{\lambda_q\,{\bf 1}_{6\rm N} & 0\cr
0 & \lambda_l\,{\bf 1}_{\rm N}},
\ee
with $\lambda_q,\,\lambda_l \in\rr_+$.
Then, rewriting the functionals (123-126) into the form 
(135-138) gives the desired relations. Clearly, the 
relation for $a_4$ simply follows by direct comparison of 
the corresponding fermionic functionals.\\

\noindent
{\bf Remark 10:} Note that we still have one
more free parameter, $\alpha$ say, to introduce in our 
model kit and to write instead of (85)
$${\cal I}_{\tilde{\rm D}} = (\psi, \dit\psi)_{\Gamma(\ep)} +
\alpha\; {\rm res}_\zeta(\dit^{-2n + 2}).$$
Of course, this additional parameter indicates that in the
definition of the universal action (2) the fermionic and the 
bosonic action functionals are considered as independently of 
each other. Alternatively, one may put $\alpha\equiv 1$ 
and then rescale both functionals independently (although
we only have one over all constant!). This is what we did, 
actually, to obtain (132) and (133-134).\\
Note that the right normalization of (124) crucially depends on 
the relations (112), which the hypercharges have to satisfy. 
Also note, the reason that each generation of quarks and each 
generation of leptons is equally weighted by the corresponding
two constants $\lambda_q$ and $\lambda_l$, respectively, is 
a consequence of the arbitrariness of the matrices 
\boldmath$\Lambda\,$\unboldmath used in the definition of 
the endomorphism $\phi'$.\\

The derived relations (144-150) can be related to 
physically measureable parameters. First, we have\\
 
\noindent
{\bf Lemma 7:} {\it The "electroweak angle" $\theta_{\rm W}$
has the range}
\bb
0.25\leq\,\sin^2\theta_{\rm W}\,\leq 0.45\,.
\ee
\noindent
{\sl Proof}: By definition, the electroweak angle 
$\theta_{\rm W}$ measures the portion of electromagnetism
to weak force:
\bb
\sin\theta_{\rm W} := 
\frac{|\!|{\bf T}_3|\!|}{|\!|{\bf Q}|\!|}.
\ee
The norm $|\!|\; .\; |\!|$, used here is defined with respect
to the "{\rm su}({\rm N}) normalization":
$\kappa({\bf E}_a, {\bf E}_b) := 
\frac{1}{2g^2_{\rm N}}\,\delta_{ab}$, where 
$\{{\bf E}_a\}_{1\leq a\leq{\rm dim}{\,\rm su}(\rm N)}$ 
denotes an appropriate basis in the fundamental 
representation  of ${\rm su}(\rm N),\, {\rm N}\geq 2$. To 
explain this more mathematically, we remind of the fact 
that the general Killing form ${\tilde\kappa}$ on a simple 
Lie algebra ${\cal G}$ may be  written as ${\tilde\kappa}(a,b)
=\frac{\lambda}{g^2}\,(\rho'(a), \rho'(b)\,)\;
\forall a,b\in{\cal G}$, where 
$( . , . )$ denotes any ad-invariant scalar product on
$\rho'({\cal G})\subset{\rm End}({\rm V})$ with 
representation ${\cal G}\mapright{\rho'}{\rm End}({\rm V})$. 
Here, the constant $\lambda$  depends on the 
scalar product and $g$ is an arbitrarily positive constant, 
parametrizing the scalar product. This holds true also when 
${\cal G}$ is semi-simple. However, the constant $g$ - 
the "coupling constant" in physical terms - may be chosen
differently for each simple component. Without loss of 
generality we can assume that
$\lambda\,(\rho'({\bf E}_a), \rho'({\bf E}_b)\,) = \delta_{ab}$
to obtain the well-known formula
\bb
\sin^2\theta_{\rm W} = 
\frac{g^2_{(1)}}{g^2_{(1)} + g^2_{(2)}}
\ee
for the electroweak angle. Hence, 
\bb
\sin^2\theta_{\rm W} =
\frac{\lambda_l+3\lambda_q}
{(1+2y_l)\lambda_l + 3(1+2y_q)\lambda_q}.
\ee
The range (155) then follows by the appropriate 
numerical values (113-114) of the hypercharges.\\

\noindent
{\bf Remark 11:} Since all norms proportional to each other 
give the same $\sin\theta_{\rm W}$  we may choose, 
alternatively, $(\rho'(a), \rho'(b)\,) := 
{\rm tr}({\bf{\rm z}}\rho'(a)^*, \rho'(b))$ to define the 
norm in (156). Hence,
\bb
\sin^2\theta_{\rm W} = 
\frac{({\bf T}_3, {\bf T}_3)}{({\bf Q}, {\bf Q})}
\ee
which, again, leads to (158). In this form, however, it becomes
evident how the range (155) does depend on the generalized 
Dirac-Yukawa Operator of the standard model, namely just 
by fixing the commutant.

The analog holds true for the ratio $g^2_{(2)}/g^2_{(3)}$
of the weak and strong coupling constants, yielding
\bb
g^2_{(2)}/g^2_{(3)} = 
\frac{4\lambda_q}{3\lambda_q + \lambda_l}.
\ee
Note that when we disregarded the possibility to introduce
the element $\zeta$ in the definition of the bosonic action
functional (74), the  relations (158) and (160)  were just a 
consequence of the fermionic representation $\rho$ used in 
the derived Yang-Mills action. Obviously, the same holds true
in the case of $\lambda_q\!=\!\lambda_l$, giving the
"GUT-preferred" numerical values:
\bb
\sin^2\theta_{\rm W}&=& 3/8,\cr
g_{(3)} &=& g_{(2)}.
\ee
On the today enery scale, however, $\lambda_q\ll\lambda_l$
seems to be preferred, c.f. \cite{Na}. Hence, 
$g_{(2)}\ll g_{(3)}$, which might be expected,
intuitively.\\

A more model specific relation may be obtaind concerning
the ratio $m_h/m_w$ of the "Higgs mass" and
the "W-boson mass" of the electroweak interaction. By
definition, having fixed a (classical) vacuum 
$\varphi_o\in\Sigma$, the masses of the gauge bosons are 
given by the quadratic form
\bb
\Omega^1(\mm,\rho_h'({\cal G}))
&\mapright{{\bf M}^2}&C^\infty(\mm)\cr
\alpha &\mapsto&\hbox{$\frac{1}{2}$}\,{\bf M}^2\!\!_{ab}\,
\ast\!(\alpha^a\wedge\ast\alpha^b),
\ee
with ${\bf M}^2\!\!_{ab}\!:=\!
\varphi_o^*\,[\rho_h'({\bf E}_a), \rho_h'({\bf E}_b)]_+\,
\varphi_o$ and $[\,.\,,\,.\,]_+$ the anti-commutator; 
$\{{\bf E}_a\}_{1\leq a\leq{\rm dim}{\cal G}}$ is a 
basis in the (semi-simple) Lie algebra ${\cal G}$ of the 
structure group G, so that 
$\alpha\!=\!\alpha^a\!\ot\!\rho_h'({\bf E}_a)$. In the
unitary gauge: $\varphi_o\!=\!(0, v/\sqrt{2})^T$ we get
$m^2_w = g^2_{(2)}v^2/4$.

The mass (matrix) of the Higgs field is defined with 
respect to the quadratic form:
\bb
\Gamma({\rm E}_h)&\mapright{{\bf M}^2_h}&
C^\infty(\mm)\cr
h &\mapsto &
\hbox{$\frac{1}{2}$}\,
(h\, ,V''|_{\varphi=\varphi_o}h)_{{\rm E}_h}, 
\ee
with $h\!:=\!\varphi\!-\! \varphi_o$ and 
$(\;.\,,\,.\,)_{{\rm E}_h}$ the induced scalar product on the 
subbundle ${{\rm E}_h}\!\subset\!{\rm E}$. In the 
unitary gauge this yields: $m^2_h = 2\lambda\,v^2$.\\

Consequently, using (147) and (149) we end up with
\bb
\frac{m_h^2}{m_w^2} =
\frac{4(2n-1)(2n-3)}{{\rm N}\pi^2}\,
\left(\frac{l}{l_p}\right)^4\,
\frac{3\lambda_q+\lambda_l}{(4\lambda_q+\lambda_l)^2}\,
(3\lambda_q\Lambda_q^4+\lambda_l\Lambda_l^4)(a_0a_2)^2,
\ee
with the abbreviation: 
$\Lambda^4\!:=\!
{\rm tr}(\hbox{\boldmath$\Lambda$\unboldmath}\!^*
\hbox{\boldmath$\Lambda$\unboldmath})^2$. Some 
further investigations similar to those in \cite{IKS2} are
needed in order to find out whether there are sufficiently 
enough relations between the unknowns on the righthand side 
of (164) determining a range where the Higgs mass has to lie 
in.

We finish this section by considering the special case of
$\hbox{\boldmath$\Lambda$\unboldmath}\equiv{\bf g}$, 
yielding the following\\

\noindent
{\bf Lemma 8:}  {\it Let $\phi' = -i\phi$ as defined by 
(110). Then the mass squared of the Higgs field 
and of the W-boson reads
\bb
m_h^2 &=& 
\frac{2(2n-1)}{(n-1)}\left(\frac{a_0}{a_3}\right)^2\,
\frac{3M_q^4 + M_l^4}{3M_q^2 + M_l^2},\\[.5em]
m_w^2 &=& 
\frac{(n-1)}{2(2n-3){\rm N}}\,\left(\frac{a_3}{a_2}\right)^2\,
\frac{3M_q^2 + M_l^2}{3\lambda_q + \lambda_l}
\ee 
with $M_q^2:=
{\rm tr}(\hbox{\boldmath$\lambda$\unboldmath}_q
{\bf m}_q^*{\bf m}_q)$ and} $M_l^2:=
{\rm tr}(\hbox{\boldmath$\lambda$\unboldmath}_l
{\bf m}_l^*{\bf m}_l)\,$\footnote{Note, by abuse of
notation $M^4\equiv
{\rm tr}(\hbox{\boldmath$\lambda$\unboldmath}
({\bf m}^*{\bf m})^2\,)$}. Here, we used the abbreviation
${\bf m}_q^*{\bf m}_q\equiv{\bf m}_d^*{\bf m}_d+
{\bf m}_u^*{\bf m}_u$ and
$\hbox{\boldmath$\lambda$\unboldmath}_q:=
\lambda_q{\bf 1}_{\rm N},\,
\hbox{\boldmath$\lambda$\unboldmath}_l =
\hbox{diag}(\lambda_{l_1},\dots,\lambda_{l_{\rm N}})
\in{\bf M}_{\rm N}(\rr)$. Correspondingly,
$\lambda_q:=
{\rm tr}\hbox{\boldmath$\lambda$\unboldmath}_q/{\rm N}$, 
as before; However, $\lambda_l:=
{\rm tr}\hbox{\boldmath$\lambda$\unboldmath}_l/{\rm N}$. 
Note that by the commutant all irreducible subspaces of the 
fermionic representation space V are now 
independently weighted. Because of the 
Kobayashi-Maskawa matrix the quark sector is considered as 
irreducible.\\

\noindent
{\sl Proof:} Since $m_h^2 = 2\lambda\,v^2$ we have with (147)  
\bb
m_h^2 = 16\alpha_0\,
\frac{3M_q^4+M_l^4}{v^2};
\ee
The relation (148) then implies (165). With 
\bb
l^2 &=& \frac{1}{2(2n-1)}\,\left(\frac{1}{a_0}\right)^2\,
\frac{3M_q^2 + M_l^2}{3M_q^4 + M_l^4},
\quad\hbox{and}\\[.5em]
v^2 &=& \frac{(n-1)}{\pi (2n-1) {\rm N}}\,
\left(\frac{a_3}{a_0}\right)^2\,
\frac{(3M_q^2 + M_l^2)^2}{3M_q^4 + M_l^4}\,
\frac{m_p^2}{4\lambda_q + \lambda_l}
\ee
an analogous calculation  yields (166).\\

Note that in the two geometrically distinguished cases where 
either the Higgs field defines a certain supercurvature or a 
certain differential form, the Higgs-form $\omega_\phi$, 
as discussed in (94-95), the Higgs mass is determined by 
the mass of the fermions, and its range depends on the 
weights of all the irreducible subspaces of the fermionic 
representation. In this sense the parameter "Higgs-mass" in 
the standard model may be considered as {\it derived} from 
the generalized Dirac-Yukawa operator of the standard model.
Clearly, the same holds for all other (possible) derivable
relations between physical parameters. We stress that the 
commutant in the definition of the bosonic 
functional (85) is motivated mainly by the fact that absolute 
values of the physical parameters, like the masses or charges 
of the particles involved, are of no physical significance. 
In this sense the generalization (74) becomes very plausible.
Again, one has to take into account that the parameters
defining the commutant are scale dependent in general.\\

\section{Outlook}

To summarize: In this paper we have introduced a particular 
geometrical model building kit based on 
the notion of generalized Dirac operators, considered as a
triple $({\rm G}, \rho, \di)$. Correspondingly, the geometrical
setting is given by a Clifford module bundle $(\ep, c)$ over a 
(closed, compact) Riemannian (spin-) manifold $(\mm, g)$ of 
even dimension $2n>2$. Within this geometrical frame we 
proposed the universal functional
\bb
{\cal I}_{\rm D}:=(\Psi, i\di\Psi)_{\Gamma(\ep)} +
{\rm res}_\zeta(\di^{-2n+2})
\ee
on $\Gamma(\ep)\!\times\!{\cal A}(\ep),\,
\di_{\nabla}\!=\!\di$, generalizing the classical action 
functional of the standard model. Indeed, we have shown how 
the action functional of the standard model - with the gravity 
action including - can be derived from a  
generalization of the Dirac-Yukawa operator. For this we
have introduced a certain class of Dirac operators, 
parametrized by some constants. Two particular choices of
these constants are distinguished geometrically. 
In particular, we have shown that the structure of the Yukawa 
coupling is such that it naturally defines a certain one form 
$\omega_\phi\!:=\!\delta_\xi\Phi$ on a twisted Clifford 
module bundle $\ep$ (the Higgs-form) and hence a particular
connection thereof. Using this connection to define a certain
Dirac operator and then evaluate the proposed functional 
(85) (with respect to diagonal sections) one obtains the full 
action of the standard model. Having derived the action we 
have shown how the free parameters of the model are linked to
physical quantities like masses and charges of the
various fields involved. Depending on the number of 
independent parameters of the model this may then lead to
non-trivial relations between the physical quantities. For
a special case we derived a relation between the mass of the
Higgs field and the masses of the fermions.

The basic idea is that the fermionic interactions determine 
the dynamics of {\it all} the fields involved in the theory. In 
the case of the standard model the fermionic interaction 
is defined with respect to the Yukawa-coupling (besides the
gauge covariant coupling), giving rise to a 
non-standard Dirac operator - the Dirac-Yukawa operator. 
Here, "non-standard" means that the appropriate class of 
connections, defining the Dirac operator in question, does not 
contain a Clifford connection as a representative and thus 
indicating that the basic geometrical setting is that of a 
Clifford module bundle instead of the particular case of a 
twisted spinor bundle. Indeed according to the Higgs form,
the tensor product structur of the Clifford module is
ignored. In contrast to mathematical applications, where 
mainly SDOs are of interest, in physics non-SDOs 
seem to play a dominant role. Let us assume that all
particles are massless. In this case the fermionic interaction 
is described by a gauge potential. Accordingly, the 
corresponding Dirac operator $\di$ is a SDO. However, 
in our scheme one has to consider not this operator but, 
instead, $\dit\!:=\!D+{\cal J}(c(F))$, where $F$ is the 
relative curvature (in this case the twisting curvature) on 
the Clifford module bundle $\ep$. Clearly, this 
new Dirac operator $\dit$ on $\tilde\ep$ is non-standard. 
Evaluation of the functional (85) with respect to this operator 
leads to the Einstein-Hilbert-Yang-Mills action and hence 
the description of the full dynamics of the physical system 
under consideration, c.f. \cite{AT2}. Therefore, although in 
the case where the dynamics of the fermions are described by 
a SDO it seems to be more natural, actually, to consider the 
non-SDO $\dit$, describing the full dynamics. As a 
consequence of this scheme, however, one neccessarly has to 
double the "internal degrees of freedom" of the fermions: 
${\rm E}\rightarrow{\tilde{\rm E}}\!:=
\!{\rm E}\!\op\!{\rm E}$. Moreover, both parts have to carry 
the same fermionic representation. Note that this doubling
may be described, geometrically, as the pullback bundle
$\triangle\!^*({\rm E}\!\times\!{\rm E})$ of 
${\rm E}\!\times\!{\rm E}\rightarrow
{\mm}\!\times\!{\mm}$ with respect to the diagonal map
${\mm}\!\ni x\mapright{\triangle}(x,x)\!\in
{\mm}\!\times\!{\mm}$. The latter may be identified with
${\mm}\!\times\!\{\pm1\}$ and thus, implicitly, also a 
doubling of spacetime is involved in our scheme. Since this 
construction is fundamental we are left to interpret this 
doubling physically (c.f. dicussion below).

With respect to the standard model, our model building 
kit somehow parallels the Connes-Lott approach. In fact,
both approaches yield analogous results concerning the
relations between the various parameters of the models. It 
therefore might be worth comparing both approaches, though
the general mathematical frame work is quite different and
there is no doubt that from a mathematical point of view,
Connes' non-commutativ geometry is much more indepth, c.f. 
\cite{Co1}, \cite{K2} and \cite{VB} for a good review. Obviously,
both kits have in common that the basic building block are 
Dirac operators. In the Connes-Lott scheme it is the internal 
Dirac operator ${\cal D}$ and in the kit proposed here it is 
the Dirac-Yukawa operator $\dipi$. As already mentioned in
the introduction the basic ideas of our model and, especially, 
the approach of considering the Yukawa-coupling as the 
fundamental input to derive the bosonic action of the standard
model with gravity including was already proposed in 
\cite{AT3}\footnote{By use of the correct Yukawa coupling (9), 
the generalized Dirac-Yukawa operator in \cite{AT3} is of
the form (95) of the paper at hand, and thus gives rise to the 
EHYMH-action}. Indeed, also in the Chamseddine-Connes 
scheme the Dirac-Yukawa operator is considered as the main 
ingredient, c.f. \cite{CC1}, \cite{IKS1}. In this scheme the 
universal action is defined by the heat trace in contrast to the 
original Connes-Lott description of the standard model, where 
the action was defined via the Dixmier trace (c.f. the 
corresponding remarks in the introduction). Concerning the 
standard model, in the Chamseddine-Connes approach one has 
to consider all Seeley-deWitt coefficients at least up to order 
four in the asymptotic expansion of the heat trace since the 
heat kernel is defined with respect to the Dirac-Yukawa 
operator $\dipi$. In our scheme, however, the generalized 
Dirac-Yukawa operator ${\dit}_\phi$ is considered as the 
basic Dirac operator. Hence, the full action of the standard
model is given by the subleading term of the asymptotic 
expansion of the corresponding heat trace. Note that the latter 
yields exactly the action of the standard model, which is not
the case in the Chamseddine-Connes scheme. Since in this
scheme the derived action obtaind by the Dirac-Yukawa 
operator is more general than the action of the standard model 
(it also contains the well-known {\it quadratic} terms in the
curvature of the base manifold), one obtains more constraints 
for the parameters involved in the model, c.f. \cite{IKS1}. 
However, the relations (112) for the weak hypercharges and 
the relations (158) and (160) for the electroweak angle and 
the coupling constants, respectively, are merely a consequence 
of the Yukawa coupling (9) and that all fields involved in the 
model carry the fermionic representation. Hence to this 
respect, the Chamseddine-Connes scheme and our kit yield 
similar results.

The power "$-2n+2$" in the definition of
the action ${\cal I}_{\rm D}$ proposed here is motivated by 
the fact that there exists a generalized Dirac operator 
(in our sense) that gives back the {\it exact} action of the 
standard model. Infact, $\sigma_2$ is the only coefficient
in the asymptotic expansion of the heat trace, which is 
linear in the curvature of the base manifold. Also by formal 
analogy between the asymptotic expansion of the heat trace and 
the asymptotic expansion of the effective action in quantum field 
theory one may expect, intuitively that the "classical action" 
is covered by the subleading term, only. As already mentioned, 
while the bosonic functional (68) was already considered in \cite{Co2} 
and in \cite{K1} in the case of the pure gravity action and in 
\cite{KW} from a somewhat more general perspective, the 
possibility to also derive the Yang-Mills and Higgs action 
from (68) was not taken into account.\\

If the former is considered as "facts" we now turn to some
"fictions". As we have already mentioned, the price we have
to pay for "adding even terms" to Dirac operators and
thereby changing the emphasis from SDOs to non-SDOs
is the additional structure $({\tilde\ep}, {\cal J})$, which
we have to introduce in our model.  Of course, this needs
some physical interpretation. A physically satisfying way to 
understand this additional structure may consist in 
interpreting the doubling of the internal degrees of fermionic 
freedoms by introducing the notion of "antiparticles" in our 
scheme. Hence, we have to incorporate the notion of 
"charge conjungation" within our model, which will be done
in a forthcoming paper where we shall also investigate in 
more detail the relation (164). 

Another point that we have in mind concerns
spectral geometry. Since the whole dynamics of the fields
involved in a physical theory should be determined by a
single (generalized) Dirac operator and, moreover, the 
corresponding fermionic action is proposed  - naturally 
enough - to take its well-known form, one may ask whether
the bosonic action actually can be considered as 
"generalizing" the fermionic action. Indeed, it is well-known
by physicists that - in a sense - the bosonic action can be
recoverd from the fermionic action by considering the former
as a "one-loop" correction of the latter. For this, one has to
introduce a zeta-regularized determinant of a certain 
operator. And this may be the point where spectral geometry 
comes in. In other words, if the bosonic action 
is considerd as a modification of the fermionic action, 
like above, we are left with the mathematical question of how 
the proposed bosonic action in our model can be 
expressed as a zeta-regularized determinant of a Dirac
operator. Moreover, as we have seen, from a geometrical point 
of view the action functional of the standard model is  but the 
subleading term of the asymptotic expansion of the heat trace 
of a certain Hamiltonian. Therefore, it might be natural to
ask whether the higher terms in the expansion permit a 
physical interpretation as well. Again, so far this is but 
fiction. However, it might be worth investigating these points 
more carefully in a future paper. 

Still quite another point, of course, is concerned with the fact 
that we are dealing with Riemannian manifolds instead of 
Lorentzian manifolds. Hence, the notion of gravity is just 
formal. However, our model may be flexible enough to work 
also in the case when Lorentzian manifolds are 
considered. The point here is that also the kernels of 
differential operators of "Huygens-type" have an asymptotic
expansion like the heat trace of an elliptic operator, c.f.
\cite{Ba}. Of course, spectral geometry in this case is not 
so well established, but see, e.g., \cite{BK} and the 
corresponding references therein. Finally, since the basic 
geometrical setting of our model kit is a Clifford module 
bundle and by the fact that there is a one-to-one 
correspondence between Dirac operators and Clifford 
superconnections, it might be possible to incorporate the 
notion of "supersymmetry" in our scheme. Thus, one may put 
more emphasis on the "superformalism", as developed by 
Quillen et. al., c.f. \cite{BGV} and \cite{HPS}, 
\cite{NS} in the case of physics. Concerning our model, this 
point of view was taken, e.g., in \cite{A}.\\

We finish this paper by a citation, which expresses the 
feeling of many physicists. We do, however, hope to have 
convinced the reader that in fact the contrary holds true:
\begin{verse}
$"\ldots$ this prescription [via the Yukawa-coupling] of the 
fermion masses is one of the least satisfactory aspects of the 
theory [standard model]. It is an entirely {\it ad hoc} 
procedure $\ldots$"\footnote{This is taken from
the instructive book {\it Gauge Theories in Particle
Physics} by I. J. R. Aitchison and A. J. Hey, 
ADAM HILGER LTD, Bristol, page 260.} 
\end{verse}

\vfill 

\noindent{\bf Acknowledgements}\\
I would like to thank Th. Sch\"ucker for explaining me the
beauty of non-commutative geometry and its application to
physics. But much more I like to thank him for his
"stubborness". Also, I like to thank Th. Ackermann for
many discussions on this subject, yielding to 
three joined works and E. Binz for his encouragements. 
Finally, I would like to thank M. Lesch for very helpful 
discussions.

\end{document}